\documentstyle[12pt]{article}
\newcommand{\be}{\begin{equation}}
\newcommand{\ee}{\end{equation}}
\newcommand{\hy}{\hat{y}}
\newcommand{\by}{\bar{y}}
\newcommand{\bz}{\bar{z}}
\newcommand{\go}{\omega}
\newcommand{\half}{\frac{1}{2}}
\newcommand{\ga}{\alpha}
\newcommand{\dga}{{\dot{\alpha}}}
\newcommand{\dgb}{{\dot{\beta}}}
\newcommand{\gb}{\beta}
\newcommand{\ws}{\wedge\star\,}
\newcommand{\bee}{\begin{eqnarray}}
\newcommand{\eee}{\end{eqnarray}}
\begin{document}
\begin{flushright}
{hep-th/9611024}\\
\vspace{1mm}
 FIAN/TD/24--96\\
\end{flushright}

\begin{center}
{\large\bf
HIGHER-SPIN GAUGE THEORIES IN FOUR, THREE\\
\vskip3mm
 AND TWO DIMENSIONS}
\vglue 1  true cm
\vskip1cm
{\bf M.A.~VASILIEV}
\vskip1cm

\medskip
I.E.Tamm Department of Theoretical Physics, Lebedev Physical Institute,\\
Leninsky prospect 53, 117924, Moscow, Russia
\medskip
\vskip2cm
Invited talk at the Quantum Gravity Seminar in memory of academician Moisei
Markov,\\ June 12-19, 1995 Moscow
\end{center}
\vskip2cm
\begin{abstract}
We review the theory of higher-spin
gauge fields in four and three space-time dimensions
and present some new results on higher-spin gauge
interactions of matter fields in two dimensions.
\end{abstract}

\newpage
\pagestyle{myheadings}

\setcounter{page}{2}

\section{Introduction}

The aim of this talk is twofold.
We review the previously obtained
results in the theory of higher-spin
gauge fields in four and three space-time dimensions
and present some new results
on higher-spin gauge interactions of matter
fields in two dimensions.

The problem of existence of  consistent gauge invariant theories
of interacting massless fields of higher spins
is one of fundamental questions in field theory.
A very stimulating argument came in the late seventies
due to the supergravity theory \cite{N} after it was realized that
the restriction $s\leq 2$
on the spins of the particles in the
massless supergravity supermultiplet leads to
a famous limitation $N\leq 8$ on the number of gravitinos
which plays a crucial role in supergravity theory
and is a direct consequence of the former restriction
exhibiting the inability to work with
interacting higher-spin gauge fields at that stage.

Another motivation comes from the superstring theory which is known
to describe infinite collections of higher-spin excitations of all
spins \cite{str}. In string theory all higher-spin excitations are massive.
One can
speculate however that they can be obtained by virtue of
a certain spontaneous
breaking  mechanism from some symmetric phase of the theory.
The
crucial question is then what is a fundamental theory of
interacting gauge higher-spin fields which underlies
this most symmetric phase of the string theory?

Theories of interacting massless fields
of all spins indeed exist \cite{FV1,more}
constituting a new class of gauge theories
based on certain infinite-dimensional gauge
symmetries, higher-spin gauge symmetries \cite{FV2,V3,KV}.
An important property of the higher-spin theories
 is that infinite-dimensional
higher-spin gauge symmetries
contain lower-spin $(s\leq 2)$ gauge
symmetries as (maximal) finite-dimensional subalgebras.
As a result
higher-spin gauge theories describe infinite
collections of higher-spin massless fields
of all spins $0\leq s<\infty $ and
 generalize naturally usual lower-spin gauge theories
containing them as subtheories (truncations).
Thus, higher-spin theories can be thought of
as most general gauge theories in $3+1$ space-time dimensions.
One can speculate that the fact that higher-spin gauge symmetries are
infinite-dimensional offers good perspectives for constructing
quantum-mechanically consistent theories unifying gravity with other
interactions
provided that higher-spin gauge symmetries are powerful enough to
ensure the cancellation of divergences.  On the other hand, that
the theory of higher-spin gauge fields in 3+1 dimensions contains
infinitely many fields of all spins makes it indeed reminiscent of
superstring theory thus giving an additional argument in favor of the
relationship of higher-spin theories with an unbroken
phase of string theory.

In fact, the infinite-dimensional
higher-spin symmetries are closely related to
(centerless)
$W_{1+\infty}$
algebra and its further generalizations described below.
So, higher-spin
gauge theories were shown
\cite{FV2,V3,KV} to be gauge theories of $W_{1+\infty}$
before the name $W_{1+\infty}$ was invented \cite{name}.
Since
$W_{1+\infty}$ is a fundamental symmetry which nowadays proves to be
important in many physical models such as conformal models, integrable
systems etc., one can speculate that a gauge theory of
$W_{1+\infty}$ may be intrinsically related to all these models.

The property that higher-spin theories describe
infinite collections of fields means
that one has to develop adequate methods
to handle them efficiently.
One of such methods we
are going to focus on in this report,
which we call ``unfolded formulation",
 allows us to formulate dynamical
equations of a system under investigation as some zero-curvature
equations supplemented with certain constraints which do not involve
any space-time derivatives. This formulation is remarkable
on its own right because it
allows one to reduce entirely the
dynamical content of the theory to the analysis of the constraints.
The crucial point here is that such a formulation requires infinitely
many auxiliary fields which appear very naturally in the higher-spin gauge
theories. In principle one can use analogous
 formulation in any other relativistic
theory that may be useful for the analysis of the
standard non-linear field theories like Yang-Mills and Einstein
theories.

Generally, the formulation of dynamical equations in the ``unfolded form"
does not imply automatically that the system is solvable because the
afore mentioned constraints may be difficult to solve themselves.
The remarkable feature of the new
1+1 model we focus on in this talk is that it does not require
 any constraints at all.
As a result, the model of higher-spin gauge interactions for matter fields
in two dimensions we present in this talk
turns out to be integrable. Let us stress that this
model is not conformal, while its integrable (in fact topological)
form is a consequence of gauging the higher-spin symmetries of d2 matter fields.

\section{Lower-Spin Examples}

The characteristic property of gauge theories
is invariance
under local symmetries, i.e. symmetries with parameters being arbitrary
functions of the space-time coordinates.

Historically, the first example of a gauge field theory was
provided by the
Maxwell theory of electromagnetism. In this case, the gauge field is identified
with the vector potential $A_\nu $ which gives rise to the field strength
$$
F_{\nu\mu}=\partial_{\nu}A_{\mu}-\partial_{\mu}A_{\nu},\qquad
\partial_{\nu}=\frac{\partial }{\partial x^{\nu} },\qquad \nu=0 \div 3
\label{f}
\eqno{(1a)}
$$
 invariant under the gauge (gradient) transformations
$$
\delta A_\nu =\partial_\nu \varepsilon
\eqno{(1b)}
$$
with an arbitrary gauge parameter $\varepsilon (x)$. As is well known, the gauge
invariant Maxwell action,
$$
S= -\frac{1}{4} \int d^4 x\,F_{\nu\mu}F^{\,\nu\mu},\qquad \delta S=0
\eqno{(1c)}
$$
describes spin-1 massless particles, photons.

 Maxwell theory can be generalized to  Yang-Mills theory
\cite{YM},
by introducing a system of mutually charged spin-1 particles described
by matrix-valu\-ed potentials $A_{\nu}{}_i {}^j$
taking values in some
Lie algebra $h$. The corresponding field strengths, gauge transformations and
action read, respectively
$$
G_{\nu\mu}=\partial_\nu A_\mu - \partial_\mu A_\nu +g\,[A_\nu ,\,A_\mu],
\eqno{(2a)}
$$
$$
\delta A_\nu=\partial_\nu  \varepsilon +g\,[A_\nu ,\varepsilon\,]\,,
\eqno{(2b)}
$$
$$
S= - \frac{1}{4}\int d^4x\, tr(G_{\nu\mu}G^{\nu\mu})\,.
\eqno{(2c)}
$$
Yang-Mills theory can be thought of as a theory of interacting massless spin$-1$
particles. In fact, under some reasonable conditions on the orders of
derivatives \cite{OP1}, the principle of gauge symmetry fixes spin$-1$ interactions
unambiguously up to a choice of a gauge group.

The second text-book example of a gauge theory is
general relativity. Here
the role of the
 gauge field is played by the metric tensor
$g_{\mu\nu}$
 while gauge transformations are identified with the
diffeomorphisms
\begin{equation}
\delta g_{\nu\mu}=\partial_\nu(\varepsilon^\rho) g_{\rho\mu}+\partial_\mu(\varepsilon^\rho )
g_{\rho
\nu} + \varepsilon^\rho \partial_\rho(g_{\nu\mu})\,,
\end{equation}
where $\varepsilon ^\rho (x) $ are infinitesimal parameters.
The gauge
principle identifies with the Einstein equivalence
principle.
The invariant Einstein-Hilbert action
\begin{equation}
S=-\frac{1}{4\kappa^2}\int \sqrt{-\mbox{det}\mid g\mid }\,(
R +\Lambda)
\end{equation}
depends on two independent coupling constants, the gravitational constant
$\kappa $ and the cosmological constant $\Lambda $. To interpret this
theory in terms
of particles, one implements the expansion procedure
$g_{\nu \mu }= \eta _{\nu \mu }+\kappa h_{\nu \mu }$ where $\eta _{\nu \mu }$
is some
fixed background
metric (flat for $\Lambda =0$ or (anti-) de Sitter for $\Lambda \neq 0$)
and $h_{\nu \mu }$ describes dynamical perturbations.
It was shown by
Fierz and Pauli \cite{FP} for the flat
case $\Lambda =0$
that the
linearized action $S$ describes free spin$-2$ massless particles, gravitons.
Once again, under certain reasonable conditions, the Einstein-Hilbert action is
the only consistent (gauge invariant) one for a selfinteracting spin$-2$
massless field \cite{OP2,BD}.

In four dimensions, the only non-trivial modification of the spin$-1$
and spin$-2$ gauge theories is supergravity \cite{N}, the theory
which, in addition to spin$-1$ and spin$-2$ gauge fields, describes spin$-3/2$
massless fields, gravitinos, which are responsible for local supersymmetry
transformations with spinorial gauge parameters $\varepsilon_{\alpha}(x) $.
A novel feature characterizing supersymmetry is that it relates interactions for
fields carrying different spins and, in particular, for bosons and fermions.
Again, pure supergravity is the only consistent theory that describes
consistent interactions of spin $-3/2$ particles.

Thus, the conventional gauge theories are based on spin$-1$ gauge fields with
scalar gauge parameters $\varepsilon (x)$, spin$-3/2$ gauge fields with spinor
gauge parameters $\varepsilon_{\alpha}(x)$ and
the spin$-2$ gauge field with the vector
gauge parameter $\varepsilon^{\rho}(x)$. Needless to say, all these theories
are of
great physical importance. The natural question then arises whether other
possibilities related to higher-spin gauge fields ($s>2$)
with highest tensors as gauge parameters do lead to
fruitful physical models too.

\section{Free Massless Higher-Spin Fields and the Interaction Problem}

The theory of free massless fields of all spins is now developed in full detail
due to the efforts
of many authors
(see e.g. \cite{F,V1} and references therein). It was found that all free
massless fields with $s\geq 1$ are Abelian gauge fields. In particular,
integer-spin massless spin$-s$ gauge fields can be described by totally
symmetric tensors $\varphi _{\nu _1\ldots\nu _s}$ subject to the double
tracelessness condition \cite{F} $\varphi ^{\rho}{}_{\rho}{}^{\eta}{}_{\eta \nu
_5\ldots\nu _s}=0$ which becomes nontrivial for $s\geq 4$. Quadratic actions
$S_s$ for free higher-spin fields can be fixed unambiguously \cite{C} (up to
an overall factor) by the requirement of gauge invariance under the Abelian
transformations

\begin{equation}
\delta\varphi_{\nu_1\ldots\nu_s}=\partial_{\{\nu_1}\varepsilon_{\nu_2\ldots \nu_{s}\}_\nu}
\end{equation}
with the parameters $\varepsilon_{\nu _1...\nu _{s-1}}$ which are rank-$(s-1)$
totally symmetric traceless tensors,
$\varepsilon^{\rho}{}_{\rho\nu _3\ldots\nu _{s-1}}=0$.
The final result is \cite{F}

\begin{eqnarray}
S_s&=&\frac{1 }{2 }(-1)^s
\int d^4 x\,\{\partial_\nu\varphi_{\mu_1\ldots\mu_s}\partial
^\nu\varphi ^{\mu_1 \ldots \mu_s}\\ &-&\frac{1 }{2
}s(s-1)\partial_\nu\varphi^\rho {}_{\rho\mu_1 \ldots
\mu_{s-2}}\partial^\nu\varphi^\sigma{}_\sigma{}^{\mu_1\ldots
\mu_{s-2}}\nonumber
+s(s-1)\partial_\nu\varphi^\rho_{\rho\mu_1\ldots\mu_{s-2}}\partial_\sigma\varphi
^{\nu\sigma\mu_1\ldots\mu_{s-2}}\\
&-&s\partial_\nu\varphi^\nu_{\mu_1\ldots\mu_{s-1}}\partial
_\rho\varphi^{\rho\mu_1\ldots\mu_{s-1}} - \frac{1 }{4
}s(s-1)(s-2)\partial_\nu\varphi^\rho{}_\rho{}^\nu{}_{\mu_1\ldots \mu
_{s-3}}\partial_\sigma\varphi^\eta{}_\eta{}^{\sigma\mu_1\ldots\mu_{s-3}}\}\,.
\nonumber \end{eqnarray}
For $s\geq 1$ this action describes spin$-s$ massless particles
which possess two independent degrees of freedom
in $d=3+1$. Quantization of this action
leads to a unitary theory free from negative-norm states. For
$s=1$ and $2$, $S_s$
reduces to the standard lower-spin actions. Fermionic higher-spin gauge
fields can be described analogously in terms of rank-$(s-1/2)$ totally
symmetric tensor-spinors $\psi _{\nu _1...\nu _{s-1/2}(\alpha )}$ ($(\alpha )$
is a spinor index) subject to the $\gamma -$tracelessness condition $\gamma
^\sigma {}_{(\alpha )}  {}^{(\beta )} \psi ^\rho{} _{\rho \sigma \nu _4...\nu
_{s-1/2}(\beta )}=0$.  This formulation is called formalism of symmetric
(spinor-) tensors.

Once the theory of free higher-spin gauge fields is
shown to be well defined,
the next nontrivial problem is how to construct consistent
interactions for hi\-gher-spin gauge fields. Consistency of a higher-spin
gauge theory demands that it should reduce to some combination of free
higher-spin systems at the linearized level with the correct signs of
the individual
actions respecting unitarity and that a number of gauge symmetries
should
be the same for free and interacting theories, $i.e$.
the interactions are allowed
to deform Abelian gauge symmetries of free theories,
as it happens in
Yang-Mills and Einstein theories,
but not to break them down.

Important indications that nontrivial higher-spin gauge theories do exist
were originally obtained in \cite{BBB,M} where it was shown that some consistent
cubic higher-spin interactions can be constructed.
These interactions however do not contain the
gravitational interaction of massless fields.
On the other hand, the problem of
 existence of the consistent gravitational interaction is
of crucial importance because of the universal role of gravity.
The analysis of
this issue carried out by several groups \cite{AD} indicated that the
attempts to introduce higher-spin-gravitational interactions encounter
serious difficulties. Technically, the reason is quite simple: in order to
introduce interaction with gravity
respecting general coordinate
invariance, one has to covariantize derivatives,
$\partial \rightarrow D=\partial -\Gamma $.
This breaks down the invariance under the higher-spin gauge
transformations because it turns out that, in order to prove invariance of the
action $S_s$, one should commute derivatives, while the commutator of
the covariant
derivatives is proportional to the Riemann tensor, $[D\ldots,\, D\ldots
]=R\ldots\,\, $.  As a result, one concludes that the gauge variation of the
covariantized action $S_s^{cov}$ has the following structure:
\begin{equation}
\delta S_s^{cov} =R_{\ldots}(\varepsilon_{\ldots}
D\varphi_{\ldots})\neq 0
\end{equation}
and that it is not clear how to
compensate these terms by any modification of the action
or/and transformation
laws.

The resolution of this problem is tricky enough.
It was shown \cite{FV1} that
consistent cubic higher-spin-gra\-vi\-tational interactions
can be constructed if one analyzes the problem in the framework
of the expansion near the (anti-)de
Sitter background. In other words, gauge invariant and
general coordinate covariant
 higher-spin-gravi\-tational interactions contain some terms proportional to the inverse
powers of the cosmological constant which diverge in the flat limit.
This result is in agreement with the conclusions of \cite{AD} where it was
implicitly assumed that one can analyze the problem in
the framework of some expansion in powers of the Riemann tensor.
The point is that such an expansion makes sense only when the Riemann tensor
is small, $i.e$. the geometry is nearly flat.

Let us stress that the nonanalyticity
of the higher-spin interactions in the cosmological constant is
a consequence of the requirement that the higher-spin gauge symmetries are
unbroken. On the other hand, higher-spin-gauge symmetries are expected
to be
broken in an appropriate physical phase
to make  all originally massless fields massive.
A value of the cosmological
constant in this physical phase is expected to be modified too. Thus, the
nonanalyticity of the higher-spin-gravitational interactions in the
cosmological constant in the symmetric phase, does not
 prevent one from building realistic models based on higher-spin gauge
theories with broken higher-spin symmetries.

\section{Geometric Formulation of Einstein Gravity}

To illustrate some of the features of the formulation of higher-spin gauge
theories described below, let us first remind the reader
relevant facts about the
``geometric formulations" of gravity.

It is well known \cite{U,MM} that gravity can be interpreted
to some extend
as a gauge theory corresponding to an appropriate space-time
symmetry algebra $g$. Vierbein $h_{\nu}{}^{a}$ and
Lorentz connection $\omega_\nu{}^{ab}$ can be identified with the
connection
$1-$forms of $g$. For example, in the four-dimensional space-time
one can chose \cite{MM} $g$ to be
the anti-de Sitter (AdS) algebra $o(3,2),$
which
gives rise to the gauge fields $A_\nu{}^{\hat a\hat b}
=-A_\nu{}^{\hat b\hat a}$
with $\hat a,\hat b=0\div 4$, and one can set $\omega_\nu{}^{ab}=A_\nu{}^{ab}$
and $h_\nu{}^{a}=\lambda^{-1}A_\nu{}^{a4}$ with $a,b=0\div 3$. The
$o(3,2)-$Yang-Mills strengths read in these terms
\be
\label{riem}
R_{\nu\mu}{}^{ab}=\partial_\nu \omega_{\mu}{}^{ab}
+\omega_{\nu}{}^a{}_c\, \omega_{\mu}{}^{cb} +\lambda^2 h_\nu{}^a h_{\mu}{}^b
-\nu\leftrightarrow \mu\,,
\ee
\be
\label{tors}
R_{\nu\mu}{}^a = \partial_\nu h_{\mu}{}^a + \omega_{\nu}{}^a{}_c h_{\mu}{}^c
- (\nu \leftrightarrow\mu)\,.
\ee
\noindent
{}From (\ref{tors}) one recognizes that
$R_{\nu\mu}{}^{a}$
has a form of the torsion tensor in the vierbein
formulation of gravity. The constraint
$R_{\nu\mu}{}^{a}=0$
expresses the Lorentz connection $\omega_{\nu}{}^{ab}$
in terms of (derivatives of) the vierbein
$h_{\nu}{}^{a}$
provided that
$h_{\nu}{}^{a}$ is a non-degenerate matrix.
Substituting
these expressions back into the Lorentz components of the field
strength
(\ref{riem}), one can make sure that, up to the cosmological-type terms
$\lambda ^2hh $\,,
$R_{\nu\mu}{}^{ab}$
coincides with the Riemann tensor in gravity.

Then one observes that the equations
$R_{\nu\mu}{}^{ab}=0$ and
$R_{\nu\mu}{}^{a}=0$
describe anti-de Sitter
space of radius $\lambda ^{-1}$.
In fact, this is the way how AdS space appears as a vacuum
solution of the higher-spin equations considered below.

A remarkable observation by MacDowell and Mansouri \cite{MM} is that
 Einstein-Hilbert action with the cosmological term can be formulated
in terms of the curvatures (\ref{riem})
in the form
\begin{equation}
\label{MM}
S^{MM}=-\frac{1}{4\kappa^2 \lambda^2}\int d^4x\,\epsilon^{\nu\mu\rho\sigma}
\epsilon^{abcd} R_{\nu\mu ,ab}\, R_{\rho\sigma ,cd}\,.
\end{equation}
Let us note that the terms proportional to $\lambda ^{-2}$ in $S^{MM}$,
which involve higher derivatives, combine into a topological term and do
not affect the equations of motion. The $\lambda -$independent term and the
term proportional to $\lambda ^2$ reduce to the scalar curvature and the
cosmological term, respectively.

Another version of this action is due to Stelle and West \cite{SW}
who observed that there is the following $so(3,2)$ covariant version
of the MacDowell-Mansouri action
\begin{equation}
S^{SW}=-\frac{1}{4\kappa^2 \lambda^2}\int_{M^4}
\epsilon^{\hat{a}\hat{b}\hat{c}\hat{d}\hat{e}}
\phi_{\hat{e}} R_{\hat{a}\hat{b}}\,
R_{\hat{c}\hat{d}}\,,
\end{equation}
where we made use of the exterior algebra formalism
considering the field strength $R$ as a 2-form,
and $\phi_{\hat{a}}$ is an additional auxiliary $o(3,2)$
vector 0-form subject to the constraint
$\phi_{\hat{a}}\phi^{\hat{a}}=1$. The MacDowell-Mansouri
formulation can be recognized as a spontaneously broken version
of the Stelle-West formulation in a particular gauge
 $\phi_a =0$ which breaks $o(3,2)$ down to $o(3,1)$.

The situation in $2+1$ gravity is
even simpler. The relevant AdS group is
$o(2,2)$=$o(2,1)\oplus o(2,1)$. The gravitational action
proposed by Witten \cite{W} is the Chern-Simons action
for this group
\be
S^W =\int_{M_3} str (w\wedge dw +\frac{2}{3}w\wedge w\wedge w )\,,
\ee
where $w$ is the $o(2,2) $ connection 2-form.

A version of the two-dimensional gravity action used in \cite{gr}
can be formulated by analogy with the
Stelle-West action as
\be
S=-\frac{1}{4\kappa^2 \lambda^2}\int_{M^4}
\epsilon^{\hat{a}\hat{b}\hat{c}}
\phi_{\hat{c}} R_{\hat{a}\hat{b}}\,,\qquad\phi_{\hat{a}}\phi^{\hat{a}}=1\,,
\ee
where $R$ is the curvature 2-form of the d2 AdS group $o(2,1)$
and $\phi^{\hat{a}}$ is a 0-form in the adjoint representation of $o(2,1)$.

A role of the space-time symmetry algebra
$o(d-1,2)$ in these examples is twofold. On the one hand,
connection 1-forms
of this algebra are identified with the dynamical fields
of the theory. On the other hand, $o(d-1,2)$
serves as the symmetry algebra of the most symmetric vacuum space.

It is then
natural to look for an appropriate generalization of this
approach which would lead to the description of the
higher-spin dynamics.
To this end it is instructive to
 use the formalism of two-component spinors
which works in the cases $d=2,3$ and $4$ due to the isomorphisms
$o(2,1)\sim sp(2)$,
$o(2,2)\sim sp(2)\oplus sp(2)$ and
$o(3,2)\sim sp(4)$.
The gravitational gauge fields now take the form
$h_{\nu}{}^{\alpha \dot \beta }$, $\omega_{\nu} ^{\alpha \beta } $ and
$\bar \omega_{\nu}{}^{\dot \alpha \dot \beta }$
in four dimensions, $h_{\alpha\beta}$ and
$\omega_{\alpha\beta}$ in three dimensions, and $h_{\pm}$
and $\omega_{+-}$
in two dimensions
(here
$\alpha ,\beta ,\ldots =1,2$
and
$\dot \alpha ,\dot \beta ,\ldots =1,2$ are spinor indices).

The key observation then is
that the generators of
$sp(2)$ and
$sp(4)$ admit
the so-called
oscillator realization. Namely,
$sp(4)$ can be realized
in terms of bilinears
\begin{equation}
L_{\alpha\beta}=\frac{1}{2} \{\hat{y}_{\alpha} ,\,\hat{y}_\beta\},\qquad
\bar{L}_{\dot{\alpha}
\dot{\beta}} =\frac{1}{2}\{\hat{\bar{y}}_{\dot{\alpha}},\,\hat{\bar{y}}_{\dot{\beta}}\},
\qquad P_{\alpha \dot{\beta}}=\hat{y}_{\alpha}\, \hat{\bar{y}}_{\dot{\beta}}
\end{equation}
constructed from mutually conjugated bosonic oscillators $\hat y_\alpha $ and
$\hat{\bar{y}}_{\dot \alpha}$
obeying the commutation relations
\begin{equation}
\label{hei}
[ \hat{y}_\alpha ,\,\hat{y}_\beta ]=2i\epsilon_{\alpha\beta},\qquad
[\hat{\bar{y}}_{\dot{\alpha}},\,\hat{\bar{y}}_{\dot{\beta}}]=2i\epsilon_{\dot{\alpha}
 \dot{\beta}}, \qquad [\hat{y}_\alpha ,\,\hat{\bar{y}}_{\dot{\beta}}]=0\,,
\end{equation}
$\epsilon_{\alpha\beta}=-\epsilon_{\beta\alpha}$,
$\epsilon_{12}=1$,
$(\hat y_{\alpha})^{\dagger}=\hat{\bar{y}}_{\dot \alpha}$.
The algebra $sp(2)$ is realized by the generators
$L_{\alpha\beta}$
($\bar{L}_{\dot{\alpha}\dot{\beta}}$ )
constructed from only undotted (dotted) indices.

Equivalently, one can say that the gravitational
fields are $1-$forms bilinear in
the auxiliary oscillator variables,
\be
\label{ads4}
\omega_\nu =
\omega_{\nu}{}^{\alpha \beta } \hat y_\alpha \hat y_\beta +
\bar{\omega}_{\nu}{}^{\dot\alpha \dot\beta } \hat{\bar y}_{\dot \alpha}
\hat{\bar y}_{\dot \beta} +
h_{\nu}{}^{\alpha \dot\beta } \hat{y}_{\alpha} \hat{\bar y}_{\dot \beta}
\,,\qquad d=3+1\,;
\ee
\be
\label{ads3}
\omega_\nu =
\omega_{\nu}{}^{\alpha \beta } \hat y_\alpha \hat y_\beta +
h_{\nu}{}^{\alpha \beta } \psi \hat{y}_{\alpha} \hat{ y}_{ \beta}
\,,\qquad d=2+1\,;
\ee
\be
\label{ads2}
\omega_\nu =
\omega_{\nu} \frac{1}{2}\{\hat y_+ \hat y_-\} +
h_{\nu}{}^{\pm } \psi \hat{y}_{\pm} \hat{ y}_{ \pm}
\,,\qquad d=1+1\,,
\ee
where we have introduced the independent involutive element
$\psi$,  $\psi^2 =1$, for the case of $d=2+1$
and use the convention that
$\alpha =\pm$ for the case of $d=1+1$.

The instructive observation then is that the action (\ref{MM})
can be generalized \cite{MM}
to the case of supergravity via extension of $sp(4)$ to the $N=1$ anti-de
Sitter superalgebra $osp(1;4)$. In terms of gauge fields, this results in
adding spin$-3/2$ gravitino fields linear in oscillators,
$\omega_{\nu}{}^\alpha  \hat y_\alpha $
and
$\bar\omega_{\nu}{}^{\dot\alpha}  \hat{\bar y}_{\dot\alpha} $
For the case of $d=2+1$ it was shown in
\cite{sd3} that the analogous extension of $sp(2)$ to $osp(1;2)$
leads to d3 supergravity.

\section{Higher-Spin Algebras and Star Product}

A natural generalization of the above construction to higher spins
consists of
allowing all powers of the spinor oscillators $\hat y$ and $\hat{\bar y}$.

Let us consider the infinite-dimensional associative
algebra $A(2)$ spanned by all polynomials of $\hat{y}_\alpha$.
Its general element $\hat{P}$ has a form
\begin{equation}
\label{el}
\hat{P}(\hat{y} )=
\sum_{n=0}^{\infty}
\frac{1}{2i\,n!}
 P^{\alpha_1\ldots\alpha_n}
\hat{y}_{\alpha_1}\ldots \hat{y}_{\alpha_n}\,.
\end{equation}
The coefficients
 $P^{\alpha_1\ldots\alpha_n}$ are supposed to be totally symmetric
in the indices $\alpha_i$  that implies Weyl ordering
of $\hat{y}_{\alpha_j}$.
$A(2)$ is called Heisenberg-Weyl algebra.
Analogously one defines the algebras $A(2n)$ with the generating
elements $\hy _\alpha $, $\alpha=1\ldots 2n$, obeying the commutation relations
\be
\label{ghei}
[\hy_\alpha ,\hy_\beta ]= 2iC_{\alpha\beta}\,,
\ee
where $C_{\alpha\beta}$ is some non-degenerate antisymmetric matrix
considered as
as a symplectic form with the
conventions
\be
C^{\alpha\gamma}C_{\alpha\beta}=\delta^\gamma_\beta\,,\qquad
A^\alpha = C^{\alpha\beta}A_\beta\,,\quad
A_\alpha = A^\beta C_{\beta\alpha} \,.
\ee

An important property of the algebra $A(2n)$ is that it admits \cite{V3}
the following unique supertrace operation
\be
\label{nstr}
str (\hat{P}_W (\hy ))
=\hat{P}_W (0)
\ee
($\hat{P}_W (\hy )$ is Weyl ordered) such that
\be
\qquad str[\hat{P}_1 ,\hat{P}_2 \}=0 \,,\qquad \forall \hat{P}_{1,2}\in
A(2n)
\ee
with the bracket $[,\}$
defined in the following way
\be
\label{scom} [\hat{P}_1 ,\hat{P}_2\}=
\hat{P}_1 \hat{P}_2 -(-1)^{\pi_1 \pi_2} \hat{P}_2 \hat{P}_1\,,
\ee
where $\pi_{1,2}$ are the ``boson-fermion" parities introduced in the standard
fashion $\hat{P}(-\hy )=(-1)^\pi \hat{P}(\hy )$.

Higher-spin algebras $shs(2n)$ are the Lie superalgebras of
(super)commutators (\ref{scom}) constructed from the associative algebras
$A(2n )$.
The supertrace (\ref{nstr}) allows one to
build invariants of $shs(2n)$ by taking supertraces of
products of its elements. The existence of the supertrace implies
that the elements with vanishing supertrace form an ideal of this
algebra. We will use the notation $shs(2n)$ for this ideal, which is
a simple algebra spanned by all
elements (\ref{el}) such that $\hat{P}_W (0)=0$.

The oscillators $\hy_\alpha$ admit the standard
differential realization. For example, for n=2
\be
\label{dif}
\hy_1 =2i\frac{\partial}{\partial z}\,,\qquad \hy_2 =z\,.
\ee
As a result,
the algebra $shs(2)$ turns out to be isomorphic to
the centerless version of $W_{1+\infty}$
\cite{name,BWV}
which is often denoted as $W_{1+\infty}$ too. Note that the supertrace
(\ref{nstr}) is defined only for this centerless part $shs(2)$
of $W_{1+\infty}$ and cannot
be extended to the full algebra of differential operators on the circle
which allows negative powers of
$z$.

For practical manipulations with the higher-spin algebras
it is convenient to use the
language of symbols of operators \cite{sym}
instead of the operator one we started with.
The idea is simple. Given $\hat{P}(\hy )\in A(2n)$ of the form (\ref{el}) one
introduces its symbol $P(y)$ which, by definition, is a function
of the commuting variables $y_\alpha $ of the same form as
$\hat{P}(\hy )$, i.e.
\begin{equation}
\label{sel}
P(y )=
\sum_{n=0}^{\infty}
\frac{1}{2i\,n!}
 P^{\alpha_1\ldots\alpha_n}
 y_{\alpha_1}\ldots y_{\alpha_n}\,,
\end{equation}
and then defines star-product $\star$ in such a way that
$P_1 \star P_2 $ be a symbol of the operator product
$\hat{P}_1 \hat{P}_2$.
For the case of the Weyl symbols under
consideration one can derive by virtue of the Campbell-Hausdorff formula
the following star-product formula
\begin{equation}
\label{nweyl}
(P\star Q)(y)=(2\pi )^{-2n} \int d^{2n} u d^{2n} v P(y+u) Q(y+v)
exp[iu_\alpha v^\alpha] \,,
\end{equation}
where $u_\alpha$ and $v_\alpha$ are integration variables.
By its definition the star-product
is associative
\begin{equation}
\label{ass}
f\star (g\star h)=(f\star g)\star h
\end{equation}
 but non-commutative.
One can easily  check with the aid of (\ref{nweyl}) that
\begin{equation}
\label{sha}
[y_\alpha ,y_\beta ]_\star =2i
C_{\alpha \beta}\,,
\end{equation}
where $[a,b]_\star = a\star b - b\star a$.
Another important property which follows from the definition
of the star-product is that given two polynomials $P(y)$ and
$Q(y)$, $\,(P\star Q) (y) $ is some polynomial too.

Now one defines the symbol version of the
connection 1-form of the algebra $shs(2n)$ as
\begin{equation}
\label{g1f}
\omega (y\mid x)=dx^\nu
\omega_\nu (y\mid x) =
\sum_{n=0}^{\infty}
\frac{1}{2i\,n!}\,
\omega^{\alpha_1\ldots\alpha_n}(x)
y_{\alpha_1}\ldots y_{\alpha_n}\,,
\end{equation}
where $x^\nu $ are space-time coordinates and the gauge field
components
$\omega^{\alpha_1\ldots\alpha_n}(x)$ are supposed to carry additional
Grassman grading for odd $n$, i.e.
the fermion fields are anticommuting
in accordance with the
standard relationship
between spin and statistics.

The curvature 2-form has the standard form
\begin{equation}
\label{cur}
R(y\mid x)=d
\omega(y\mid x) +
\omega(y\mid x) \wedge\star\,
\omega(y\mid x)\,,
\end{equation}
where $d=dx^\nu (\partial /\partial x^\nu )$ is
the space-time exterior differential.
Note that the
second term on the $r.h.s$. of (\ref{cur}) does not vanish because of
the noncommutativity
of the star-product and automatically contains the
supercommutators (\ref{scom})
due to the anticommutativity of the fermionic gauge fields.
One can expand the curvature $2-$form (15) in powers of the auxiliary
variables $ y$
\begin{equation}
R(y\mid x)=
\sum_{n=0}^{\infty}
\frac{1}{2i\,n!}
y_{\alpha_1}\ldots y_{\alpha_n}
R^{\alpha_1\ldots\alpha_n}\,.
\end{equation}
The explicit form of the coefficients $R^{\alpha_1\ldots\alpha_n}$
can be obtained by virtue of the formula (\ref{nweyl}). We do not
need it in this talk however and refer the reader
for more details
to \cite{V3}.
Let us note that originally a form of these
curvatures has been derived from the detailed analysis of the
higher-spin dynamics \cite{FV2} while the operator realization of the
higher-spin algebras described above was found afterwards \cite{V3}.

\section{Higher-Spin Action in 2+1 Dimensions}

The algebra
of global higher-spin symmetries  of
the 2+1 problem
is $shs(2)\oplus shs(2)$ which is a sum of two simple algebras
by analogy to the case of pure gravity
with $o(2,2)\sim sp(2)\oplus sp(2)$.
Such a doubling can be introduced with the aid of the additional generating
element $\psi$ such that $\psi^2 =1$. A general element of this algebra
then has a form
\begin{equation}
\label{el3}
P(\hat{y},\psi )=
\sum_{
\begin{array}{c}
\scriptstyle{n=0}\\
\raise 1,3ex\hbox{$\scriptstyle{A=0,1}$}
\end{array}
}^{\infty}
\frac{1}{2i\,n!}
 P^{A\,\alpha_1\ldots\alpha_n}
(\psi )^A y_{\alpha_1}\ldots y_{\alpha_n}\,.
\end{equation}

A higher-spin
counterpart of the Witten gravity action
was introduced by Blencowe \cite{blen} as the Chern-Simons
action for $shs(2)\oplus shs(2)$,
\be
\label{S3}
S^{2+1}=\int_{M_3}str (\go\ws d\go +\frac{2}{3}\omega\ws\go\ws\go )
\ee
(with the convention that $str (\psi P(y))=0$). Analogously to
the case of pure gravity, in $2+1$ dimensions higher-spin gauge fields
are not propagating and do not carry their own degrees of freedom.
The dynamical equations which follow from the action (\ref{S3}) have a
standard form $R=0$ and admit only trivial solutions in the topologically
trivial situation. An interesting question then is how to introduce
interactions of higher-spin gauge fields with matter fields?
In section 10 we
describe a model which solves this problem on the level of
equations of motion.

\section{Higher-Spin Action in 3+1 Dimensions}

Higher-spin gauge fields become propagating
for $d\geq 4$. The simplest non-trivial case therefore is
$3+1\,$-dimensional space-time. For simplicity let us focus on the
purely bosonic version of the higher-spin superalgebra in $d=3+1$
which is the even (bosonic) subalgebra $hs(4)$ of $shs(4)$.
It is convenient to use the two-component spinor notations
splitting
the full $sp(4)$ spinor into the complex $sp(2)$ two-component spinor
$y_\ga$ and its complex conjugate $\bar{y}_{\dot{\ga}}$.
The algebra $hs(4)$ is thus spanned by
even power polynomials in $ y$ and
${\bar y}$, $P(y,\bar{y} )$=$P(-y,-\bar{y} )$.
The gauge fields
of
$hs(4)$ are described by the generating function
\begin{equation}
\label{gen31}
\omega ({y},{\bar{y}}\mid x)=
\sum_{
\begin{array}{c}
\scriptstyle{n,m=0;}\\
\raise 1,3ex\hbox{$\scriptstyle{n+m\,- \, even}$}
\end{array}
}^{\infty}
\frac{1}{2i\,n!m!}
{y}_{\alpha_1}\ldots {y}_{\alpha_n}{\bar{y}}_{\dot{\beta}_1}\ldots
{\bar{y}}_{\dot{\beta}_m}
\omega^{\alpha_1\ldots\alpha_n}{}_,{}^{\dot{\beta}_1
\ldots\dot{\beta}_m}(x)\,,
\end{equation}
where the  multispinor space-time
1-form coefficients
$\omega^{\alpha_1\ldots\alpha_n}{}_,{}^{\dot{\beta}_1
\ldots\dot{\beta}_m}(x)$
are identified with the physical higher-spin fields. It was
argued
in section 3
that the  massless spin$-2$ field is described by
$\omega ^{\alpha _1...\alpha _n},^{\dot \beta _1...\dot \beta _m}$ with
$n+m=2$. This relation is generalized
to an arbitrary spin $s$ as follows \cite{V2}
\begin{equation}
\label{ns}
n+m=2(s-1)\,,
\end{equation}
$i.e$.
a spin$-s$ massless particle is described with the aid of the collection
of all $1-$forms
$\omega_{\nu}{}^{\alpha _1...\alpha _n},^{\dot \beta _1...\dot \beta _m}$
with the
overall number of spinor indices fixed according to (\ref{ns}).
It is worth mentioning that
these collections of gauge fields form the irreducible
rank - $2(s-1)$ tensor representations
with respect to the
 adjoint action
of the full anti-de Sitter algebra $sp(4)$ (\ref{ads4}).

The fact that such a set of fields
properly describes spin-$s$ massless fields
 is a consequence of the explicit analysis of the higher-spin dynamics
\cite{V2} based on the
following action principle
\begin{equation}
\label{act31}
S=-\frac{1}{4\kappa^2 \lambda^2}
\sum_{
\begin{array}{c}
\scriptstyle{n,m=0;}\\
\raise 1,3ex\hbox{$\scriptstyle{n+m\, -\, even}$}
\end{array}
}^{\infty}
\frac{i^{n+m-1}}{n!m!} \epsilon
(n-m) \int_{M^4}  R_{\alpha_1\ldots\alpha_n ,\,\dot{\beta}_1\ldots\dot{\beta}_m} \wedge
R^{\alpha_1\ldots\alpha_n}{}_,{}^{\dot{\beta}_1\ldots\dot{\beta_m}}\,,
\end{equation}
where
$
R^{\alpha_1\ldots\alpha_n}{}_,{}^{\dot{\beta}_1\ldots\dot{\beta_m}}$
are the components of the full $hs(4)$ curvature tensor (\ref{cur}),
\begin{equation}
\label{cur31}
R(y,\bar{y}\mid x)=
\sum_{
\begin{array}{c}
\scriptstyle{n,m=0;}\\
\raise 1,3ex\hbox{$\scriptstyle{n+m - even}$}
\end{array}
}^{\infty}
\frac{1}{2i\,n!m!}
{y}_{\alpha_1}\ldots {y}_{\alpha_n}{\bar{y}}_{\dot{\beta}_1}\ldots
{\bar{y}}_{\dot{\beta}_m
} R^{\alpha_1\ldots\alpha_n}{}_,{}^{\dot{\beta}_1
\ldots\dot{\beta}_m}(x)\,.
\end{equation}

The explicit analysis of the
quadratic part of the action (\ref{act31}) shows
\cite{V2} that
its variation
with respect to the
``extra fields",
$\omega_{\alpha_1 \ldots\alpha_n\,,\beta_1 \ldots\beta_m}$ with $|n-m|>2$,
vanishes
identically, while the variation with respect to the dynamical fields,
$\omega_{\alpha_1 \ldots\alpha_n\,,\beta_1 \ldots\beta_m}$ with $|n-m|\leq 2$,
is non-trivial
and leads to
the correct free equations
for massless fields.

Since extra fields contribute
to the interaction terms one has to express them in terms
the dynamical fields to have a well defined non-linear action.
The appropriate constraints have the form \cite{V2}
\be
\label{con1}
h_{\{\alpha} {}^{\dot{\gamma}}\wedge R_{\alpha_1 \ldots\alpha_n\}_\alpha \,,
\dot{\beta_1}\ldots\dot{\beta_m}\dot{\gamma}}=0\qquad n> m \geq 0\,,
\ee
\be
\label{con2}
h^{\gamma}{}_{\{\dot{\beta}}\wedge R_{\alpha_1 \ldots\alpha_n \gamma \,,
\dot{\beta_1}\ldots\dot{\beta_m}\}_\beta }=0\qquad m> n \geq 0\,.
\ee
These constraints express successively all  ``extra fields"
$\omega_{\alpha_1 \ldots\alpha_n\,,\beta_1 \ldots\beta_m}$ with $|n-m|>2$
in terms of the dynamical fields
$\omega_{\alpha_1 \ldots\alpha_n\,,\beta_1 \ldots\beta_m}$ with $|n-m|\leq 2$.
It is important that these  constraints
are algebraic with respect to the extra fields expressing the latter
in terms of derivatives of the dynamical fields.
The constraints (\ref{con1}) and
(\ref{con2}) play a crucial role in the description of the
higher-spin dynamics,
governing a form of the interactions of the
dynamical higher--spin fields in the action (\ref{act31}).
A consequence of this mechanism is
that higher-spin interactions for the dynamical fields
contain higher derivatives.
The same mechanism leads to the non-analyticity
of the interaction terms in
the cosmological constant.

The action (\ref{act31}) possesses the following basic properties:
%\begin{enumerate}

\noindent $(i)$
%\item
$S$ is explicitly general coordinate invariant due to the
exterior algebra formalism;

\noindent $(ii)$
%\item
in the spin$-2$ sector $(n+m=2)$ $S$ reduces to the Einstein-Hilbert
action in the MacDowell-Mansouri form (\ref{MM});

\noindent $(iii)$
%\item
on the linearized level, $S$ amounts to the sum of
free actions for all massless bosonic fields with $s\geq 2$ in the formalism of
nonsymmetric tensors \cite{V1,V2} which is dynamically equivalent to the
formalism of symmetric tensors sketched in Introduction;

\noindent $(iv)$
%\item
$S$ is gauge invariant in the cubic
order provided that the constrains (\ref{con1}) and (\ref{con2}) are imposed
\cite{FV1}.
%\end{enumerate}

Thus, the action $S$ supplemented with the constraints
(\ref{con1}), (\ref{con2}) solves the
higher-spin problem in the lowest order in interactions \cite{FV1}.
A non-trivial problem which still remains unsolved
is how to generalize
this result to highest orders in interactions.
To construct a closed action
one has to develop a formalism based on
appropriate generating functions of auxiliary spinor variables.
This is expected to lead to
a certain higher-spin generalization of the Stelle-West formulation of gravity
which
requires an appropriate counterpart of the auxiliary field $\phi^a$.  At the
moment this problem is not solved at the action level.
However an analogous program is
completed for the equations of motion.
For this reason in the sequel we
focus mainly on the formulation of equations of motion.

\section{Unfolded Formulation}

Before going into details of the full formulation of the
higher-spin equations let us discuss some general features
of the ``unfolded formulation" \cite{un} we are
going to implement and give some simple examples.

In the context of applications to the higher-spin problem
we will use a particular case
of unfolded formulation where the dynamics is described
in terms of a set of 1-forms
$\go (x)=$$dx^\nu \go_\nu^i (x) T_i $
taking values in some Lie superalgebra $l$ ($T_i \in l)$
and a set of 0-forms $B^A (x)$ which takes values in
a representation space of
some representation $(t_i ){}^B {}_A$ of $l$. The dynamical equations
are then formulated in the form
\be
\label{0cur}
d\go =\go\wedge\go\,,
\ee
\be
\label{ccov}
dB^A =\go^i t_i \, {}^A{}_B B^B
\ee
and
\be
\label{con}
\chi (B) =0 \,,
\ee
where $\chi (B)$ are some constrains which do not contain the
space-time differential $d=dx^\nu \frac{\partial}{\partial x^\nu }$
and are invariant under the gauge transformations
\be
\label{g1}
\delta {\go}=d\epsilon - [\go ,\epsilon ]\,,
\ee
\be
\label{g0}
\delta {B^A }= \epsilon^i (t_i ){}^A {}_B B^B
\ee
that  guarantees the invariance of the full system of equations
(\ref{0cur})-(\ref{con}) under the transformations (\ref{g1}) and (\ref{g0}).

Dynamical content of the equations
(\ref{0cur})-(\ref{con})
is screened in the constraints (\ref{con}).
Indeed, locally one can integrate out explicitly the first two
equations to a pure gauge solution
\be
\go=d (g^{-1} (x))g (x)\,,
\ee
\be
\label{bsol}
B(x)=t_{g(x)} (B_0 )\,,
\ee
where $B_0$ is an arbitrary $x$ - independent quantity and
$t_{g(x)}$ is the exponential of the representation $t$ of $l$.
Since the constraints $\chi (B)$ are gauge invariant one is left with the
only condition
\be
\label{con0}
\chi (B_0 )=0\,.
\ee

Suppose that $g(x_0 )=I$ for some space-time point $x_0$.
{}From (\ref{bsol}) it follows then that $B_0$=$B(x_0 )$. One can wonder
how any restrictions imposed on
values of some 0-forms in a fixed point of space-time can lead
to a non-trivial dynamics. The answer is that this is possible
if the set of 0-forms $B$ is reach enough to describe
all space-time derivatives of the dynamical fields in a fixed point of
space-time provided that the constraints (\ref{con}) just single out those
values of the derivatives which are compatible with the dynamical
equations of the system under consideration. By knowing any solution
of (\ref{bsol}) one knows all derivatives of the dynamical fields
 compatible with the field equations and can therefore reconstruct
these fields by analyticity in some neighborhood of $x_0$.
The crucial point here is that
in order to proceed along these lines one necessarily has to use
some infinite-dimensional representation $t$ for 0-forms.
{}From this point of view the special feature of
higher-spin theories is that they contain infinite collections
of dynamical fields from the very beginning
so that it is natural to introduce
infinitely many auxiliary 0-forms in these theories.

Let us now illustrate how this mechanism works for the simplest
field-theoretical model of free
massless spin-0 equations in the flat space--time
 of arbitrary dimension $d$. In this example $l$ is identified with
the Poincare algebra $iso(d-1,1)$
\be
\go_\nu =(h_\nu{}^a ,\go_\nu {}^{ab} )
\ee
($a,b =0-(d-1)$).
The vanishing curvature conditions of $iso(d-1,1)$
\be
\label{nR0}
R_{\nu\mu}{}^a =0\,,\qquad R_{\nu\mu}{}^{ab}=0
\ee
then imply that the vierbein $h_\nu {}^a$
and Lorentz connection $\go_\nu {}^{ab}$ describe the flat geometry.
Fixing the local Poincare gauge transformations one can set
\be
\label{nva}
h_\nu {}^a =\delta _\nu^a \,,\qquad \go_\nu{}^{ab}=0.
\ee
Let us note that the ambiguity in local Poincare gauge
transformations is equivalent to the general coordinate transformations
provided that  the zero-curvature conditions (\ref{nR0}) are true and the
vierbein $h_\nu {}^a$ is invertible.
As a result, the gauge fixing (\ref{nva}) is equivalent to
choosing the  Cartesian coordinate frame.

To describe dynamics of the spin zero massless field
$\phi (x)$ let us introduce the
infinite collection of 0-forms $\phi_{a_1\ldots a_n}(x)$
which are totally symmetric traceless tensors
\be
\label{tr}
\eta^{bc}\phi_{bca_3\ldots a_n}=0\,,
\ee
where $\eta^{bc}$ is the flat Minkowski metrics.
The ``unfolded" version of the Klein-Gordon equation
has a form of the following infinite chain of equations
\be
\label{un0}
\partial_\nu \phi_{a_1\ldots a_n }(x) =h_\nu {}^b
\phi_{a_1 \ldots a_n b}(x)\,,
\ee
where we have used the opportunity to replace the Lorentz covariant
derivative by the ordinary flat derivative $\partial_\nu$ due to the
flatness condition (\ref{nva}) (in any other gauge one has to replace
the flat derivative $\partial$ by the Lorentz covariant derivative).
The condition (\ref{tr}) is a specific realization of the
constraints (\ref{con}) while the system of equations
(\ref{un0}) is a particular realization of the equations (\ref{ccov}).
It is easy to see that this system is formally
consistent, i.e. the repeated $\partial_\mu$ differentiation of (\ref{un0})
does not lead to any new conditions after antisymmetrization
$\nu\leftrightarrow\mu$. This property is equivalent to the fact that
the set of zero forms $\phi_{a_1 \ldots a_n}$ spans some
representation of the Poincare algebra.

To show that this system of equations is indeed equivalent to the
free massless field equation $\Box \phi (x)=0$ let us identify the
scalar field $\phi (x)$ with the member of the family
of 0-forms $\phi_{a_1 \ldots a_n}$ at $n=0$.
Then the first two members of the
system (\ref{un0}) read
\be
\partial_\nu \phi =\phi_\nu \,,
\ee
\be
\partial_\nu \phi_\mu= \phi_{\mu\nu}\,,
\ee
where we have identified the world and tangent indices taking into account
the gauge condition (\ref{nva}).

The first of these equations just tells us that
$\phi_\nu$ is a first derivative of $\phi$.
The second one tells us that
$\phi_{\nu\mu}$ is a second derivative of $\phi$. However, because of the
tracelessness condition (\ref{tr}) it imposes the Klein-Gordon equation
$\Box\phi =0$.
It is easy to see that all other equations in (\ref{un0}) express highest
tensors in terms of the higher-order derivatives
\be
\label{hder}
\phi_{\nu_1 \ldots \nu_n}= \partial_{\nu_1}\ldots\partial_{\nu_n}\phi
\ee
and impose no new conditions on $\phi$. The tracelessness conditions
(\ref{tr}) are all satisfied once the Klein-Gordon equation is true.

Let us note that the system (\ref{un0}) without the constraints (\ref{tr})
remains formally consistent but is dynamically empty just expressing all
highest tensors in terms of derivatives of $\phi$ according to (\ref{hder}).
This simple example
illustrates how constraints can be equivalent to the dynamical equations.

The above consideration
can be simplified further by means of
introducing the auxiliary
coordinate $u^a$ and the generation function
\be
\Phi (x,u)=\sum_{n=0}^\infty \frac{1}{n\,!}\phi_{a_1 \ldots a_n}(x)
u^{a_1} \ldots u^{a_n}
\ee
with the convention that
\be
\Phi (x,0)=\phi(x)\,.
\ee
This generating function accounts for all tensors
$\phi^{a_1 \ldots a_n}$ provided that the tracelessness condition is
imposed which in these terms implies that
\be
\label{ubox}
\Box_u \Phi (x,u)\equiv \frac{\partial}{\partial u^a}
\frac{\partial}{\partial u_a} \Phi =0\,.
\ee
The equations (\ref{un0}) then acquire the simple form
\be
\label{xu}
\frac{\partial}{\partial x^\nu} \Phi (x,u)=\frac{\partial}{\partial u^\nu}
\Phi (x,u)\,.
\ee
{}From this realization one concludes, first, that
the translation generators in the infinite-dimensional
representation of the Poincare algebra formed by the higher tensors
$\phi^{a_1 \ldots a_n}$
are realized as translations in the $u$--space and, second, that
one can indeed find a general solution of the equation (\ref{xu}) in the
form
\be
\Phi (x,u )=\Phi (x+u,0) =\Phi (0,x+u )\,
\ee
from which it follows in particular that
\be
\label{tay}
\phi (x)=\Phi (0,x)=\sum_{n=0}^\infty \frac{1}{n!}
\phi_{\nu_1\ldots\nu_n }(0) x^{\nu_1} \ldots x^{\nu_n}\,.
\ee
{}From (\ref{tr}) and (\ref{hder}) one can see that
this is indeed the Taylor expansion for any solution of the
Klein-Gordon equation which is analytic in $x=0$.
Moreover one can recognize the equation (\ref{tay}) as a particular
realization of the pure gauge solution (\ref{bsol}) with
the gauge function $g(x)$ of the form
\be
g(x)=exp (x^\nu h_\nu {}^a \frac{\partial}{\partial u^a})\,.
\ee

The example of the scalar field considered here is so simple
that it tends to be trivial after introducing the auxiliary
variables $u^a$.
Remarkably, a proper generalization
of this approach to non-trivial higher-spin dynamics
is at the moment the only known one working for non-linear
higher-spin equations. Let us note that the described formalism
has some similarities with the non-linear realization approach
developed for the Yang-Mills case in \cite{e} where the relevance of the
bilocal fields analogous to $\Phi (x,u)$ subject to the equations
analogous to (\ref{xu}) was demonstrated.

\section{Higher-Spin Equations of Motion in 3+1 Dimensions}

Let us now  explain how one can formulate non-linear
higher-spin equations in $3+1$ dimensions in the unfolded form.
For simplicity we confine ourselves to the bosonic
case. A general treatment which allows one to include fermions
can be found in \cite{more}.

To describe on-mass-shell higher-spin dynamics in $d=3+1$,
we introduce \cite{more} the following set of generating functions
\be
\label{nng1}
W(Z;Y\mid x)=dx^\nu W_\nu (Z;Y\mid x)\,,
\ee
\be
\label{g2}
B(Z;Y\mid x)\,,
\ee
\be
\label{g3}
s(Z;Y\mid x)=dz^\alpha s_\alpha (Z;Y\mid x)+d\bar{z}^{\dot{\alpha}}
\bar{s}_{\dot{\alpha}}(Z;Y\mid x)\,,
\ee
where
$Z=(z_{\alpha},\, \bar z_{\dot \alpha})$ and
$Y=(y_{\alpha},\, \bar y_{\dot\alpha})$
are two independent sets of auxiliary spinor variables while $x$
denotes space-time coordinates as before. A physical meaning of the
generating functions (\ref{nng1})-(\ref{g3}) is as follows.
The space-time $1-$form $W$ is the generating function for higher-spin gauge
potentials. The 0-form $B$ serves as a generating function for
lower-spin fields (i.e., a spin$-0$ scalar) and for on-mass-shell
nontrivial higher-spin curvatures generalizing the gravitational
Weyl-tensor (see below). The space-time 0-form $s$ can be interpreted as a
$1-$form with respect to auxiliary anticommuting spinor differentials $dz^\alpha
$ and $d\bar z^{\dot \alpha }$,
\begin{equation}
\{dz^\alpha ,\,dz^\beta \} =0,\qquad \{d\bar{z}^{\dot{\alpha}}
,\,d\bar{z}^{\dot{\beta}}\}=0, \qquad\{dz^\alpha ,\,d\bar{z}^{\dot{\beta}} \}
=0\,,
\end{equation}
which commute with all other variables.
The field $s$ is auxiliary in nature, describing no independent degrees of
freedom. It serves as a differential operator shifting along the auxiliary
spinorial variables $Z$. That we confine ourselves to the purely
bosonic theory in this section means that $G(-dZ;\, -Z;\, -Y) = G(dZ;\, Z;\,Y)$
for $G=W$, $B$ and $s$.

To formulate higher-spin equations of motion, we endow the linear space of
functions $f(Z;Y)$ with a structure of associative algebra with the
$*$ product law,
\begin{equation}
\label{star}
(f*g)(Z;Y)=(2\pi )^{-4}\int d^4 Ud^4 V\, f(Z+U;Y+U)\,g(Z-V;Y+V)\,\exp
i(u_\alpha v^\alpha +\bar{u}_{\dot{\alpha}} \bar{v}^{\dot{\alpha}})\,\,,
\end{equation}
where $U=(u_\ga ,\bar{u}_{\dot{\ga}})$ and  $V=(v_\ga ,\bar{v}_{\dot{\ga}})$
are integration variables.  This product law is some
particular symbol version of the Heisenberg--Weyl algebra,
\be
\label{hhei} [y_\ga ,y_\gb ]_* =-[z_\ga ,z_\gb ]_* =2i
\epsilon_{\ga\gb}\,,\qquad [\by_\dga ,\by_\dgb ]_* =-[\bz_\dga ,\bz_\dgb ]_* =2i
\epsilon_{\dga\dgb}\,\ee
(all other commutators vanish).
The product law (\ref{star}) is associative, $(f*g)*h=f*(g*h)$, and
regular:  given two polynomials $f(Z;Y)$ and $g(Z;Y)$, $(f*g)(Z;\,
Y)$ is some polynomial too.  The latter property guarantees that the
formulae containing star products make sense for the coefficients of the power
series expansions of the generating functions $W$, $B$ and $s$.

The totally consistent system of higher-spin equations reads \cite{more}
\be
\label{beq}
dW=W*\wedge\, W \,,
\ee
\be
\label{dB}
dB=W*B-B*\tilde{W} \,,
\ee
\be
\label{ds}
ds=W*s-s*W \,,
\ee
\be
\label{sB}
s*B=B*\tilde{s} \,,
\ee
\be
\label{eeq}
s* s=dz^\alpha  dz_\alpha\, (i+B*\kappa )+d\bar{z}^{\dot{\alpha}}
d\bar{z}_{\dot{\alpha}}\, (i+B*\bar{\kappa})\,,
\ee
where
\begin{equation}
\kappa =\exp (iz_\alpha y^\alpha ),\qquad \bar{\kappa} =\exp (i\bar{z}_{\dot{\alpha}}
\bar{y}^{\dot{\alpha}})
\end{equation}
and $\sim$  changes a sign of all undotted spinors,
\begin{equation}
\tilde{f}(dz,d\bar{z};z,\bar{z};y,\bar{y})
=f(-dz,d\bar{z};-z,\bar{z};-y,\bar{y})\,.
\end{equation}

The system of equations (\ref{beq})-(\ref{eeq}) has ``unfolded form".
The equation (\ref{beq}) is
a particular case of the equation (\ref{0cur}), the equations (\ref{dB})
and (\ref{ds}) have the form (\ref{ccov}), and the equations (\ref{sB})
and (\ref{eeq}) serve as some constraints (\ref{con}). It is important
that these constraints are gauge invariant so that
the equations (\ref{beq})-(\ref{eeq}) are explicitly
invariant under the higher-spin gauge transformations
\be
\label{deltaw}
\delta W =d\varepsilon -W*\varepsilon +\varepsilon *W,
\ee
\be
\label{deltaB}
\delta B=\varepsilon *B-B*\tilde{\varepsilon},
\ee
\be
\label{deltas}
\delta s=\varepsilon *s-s*\varepsilon.
\ee
Also this system of equations is explicitly
general
coordinate covariant due to the exterior algebra formalism.

What is less straightforward to see is that the system of equations
(\ref{beq})-(\ref{eeq}) indeed describes the dynamics of massless
higher-spin fields.
The detailed analysis of this issue is presented in
\cite{more,V5} (in \cite{more} also a more general form of
higher-spin interactions has been considered).
Here, we only outline the main steps and basic ideas.

The crucial point is to show that the equations (\ref{beq})-(\ref{eeq})
describe correct free field dynamics at the linearized level.
The relevant perturbative procedure consists of
the order by order analysis
of the equations (\ref{beq})-(\ref{eeq})
in the framework of the expansion in powers of $B$ which
physically is equivalent to the expansion in powers of higher-spin curvatures
generalizing the Weyl tensor. One starts with the following ansatz for $s$
\begin{equation}
\label{lin}
s=(dz^\alpha z_\alpha +d\bar{z}^{\dot{\alpha}} \bar{z}_{\dot{\alpha}} )+ O(B)\,,
\end{equation}
which can be easily verified to solve (\ref{eeq}) in the lowest order in B.
{}From (\ref{lin})
and (\ref{hhei}), it follows that
\begin{equation}
s*f-f*s=-i\,(dz^\alpha \frac{\partial}{\partial z^\alpha} +d\bar{z}^{\dot{\alpha}}
\frac{\partial}{\partial \bar{z}^{\dot{\alpha}}})\, f(Z;Y)+O(B)f\,.
\end{equation}
As a result, the equations (\ref{ds})-(\ref{eeq}) reduce to some differential
equations with respect to $\partial /\partial z^\alpha $ and $\partial /\partial
\bar z^{\dot \alpha }$, which determine $W(Z;Y),B(Z;Y)$ and $s(Z;Y)$ itself in
terms, of the initial data $\omega (Y)=W(0;Y)$ and $C(Y)=B(0;Y)$ up to some pure
gauge transformations (\ref{deltaw})-(\ref{deltas}). In fact, these initial
data serve as the generating functions for
physical higher-spin fields. In particular, $\omega (Y)$ can
be identified with the higher-spin generating function (\ref{gen31}).
The doubling of spinor variables, $Y\rightarrow (Z,Y)$,
serves as a sort of a technical trick which
enables one to describe complicated expressions as solutions of certain
simple nonlinear differential equations with respect to $Z$.

At the second stage, inserting the expressions for $W$, $B$ and $s$
in terms of the initial data $\omega $ and $C$ back into the equations
(\ref{beq}) and (\ref{dB}), one gets
the dynamical equations for physical fields of all spins
provided that the background gravitational field is introduced as
a vacuum value $\omega_0 (y |x)$ of $\omega (y |x)$ such that the
zero-curvature equation (\ref{beq}) is true for $\omega_0 (Y |x)$.
The AdS geometry then arises as a solution with $\omega_0 (Y |x)$
of the form (\ref{ads4}).
The form of the resulting dynamical
equations is analogous to that of the spin 0 example considered in section 8.
The only distinctions are that now we use
the formalism of two-component spinors and the
analysis is carried out in the anti-de Sitter background.

In this formalism the spin-0 matter field is described by
the infinite chain of $0-$forms
$C_{\alpha_1\ldots\alpha_n,\dot \beta_1\ldots\dot\beta_n}$
which are totally symmetric multispinors. This set is
the two-component spinor version of
the set of totally symmetric traceless tensors considered in
section 8.
The equations which follow from (\ref{dB}) in this sector read
\begin{equation}
D^LC_{\alpha_1\ldots\alpha_n,\,\dot{\beta}_1\ldots\dot{\beta}_n}=
h^{\gamma\dot{\delta}}C_{\alpha_1\ldots\alpha_n\gamma,\,\dot{\beta}_1\ldots\dot{\beta}_n
\dot{\delta}}
- n^2\lambda^2 h_{\{\alpha_1\dot{\beta}_1}
C_{\alpha_2\ldots\alpha_n,\,\dot{\beta}_2
\ldots\dot{\beta}_n\}_{\alpha,\dot{\beta}}}\,.
\end{equation}
This system of equations can be shown to be equivalent to the spin 0 massless
equations for the field $C$ in the anti-de Sitter space analogously to
the flat space example considered in
section 8.
Again, the infinite chain of $0-$forms
$C_{\alpha_1\ldots\alpha_n,\dot \beta_1\ldots\dot\beta_n}$
with $n\geq 0$
describes all on-mass-shell nontrivial combinations of the
derivatives of the scalar field $C.$

Analogous analysis shows that the fields
$C_{\alpha_1\ldots\alpha_n,\dot \beta_1\ldots\dot\beta_m}$ with $|n-m|=2s$
describe massless fields of spin $s$. Let us illustrate this for the
particular case of Einstein gravity, i.e. $s=2$.
As argued in section 2, Lorentz connection $1-$forms
$\omega _{\alpha \beta }$,
$\bar \omega_{\dot \alpha\dot \beta}$
and vierbein $1-$forms
$h_{\alpha\dot \beta}$
can be identified with the $sp(4)-$gauge fields. The corresponding
$sp(4)-$curvatures read in terms of two-component spinors
\be
\label{nR}
R_{\alpha_1 \alpha_2}=d\omega_{\alpha_1 \alpha_2} +\omega_{\alpha_1}{}^\gamma
\wedge \omega_{\alpha_2 \gamma} +\lambda^2\, h_{\alpha_1}{}^{\dot{\delta}}\wedge
h_{\alpha_2 \dot{\delta}}\,,
\ee
\be
\label{nbR}
\bar{R}_{\dot{\alpha}_1 \dot{\alpha}_2}=d\bar{\omega}_{\dot{\alpha}_1
\dot{\alpha}_2} +\bar{\omega}_{\dot{\alpha}_1}{}^{\dot{\gamma}}
\wedge \bar{\omega}_{\dot{\alpha}_2 \dot{\gamma}} +\lambda^2\,
h^\gamma{}_{\dot{\alpha}_1} \wedge h_{\gamma \dot{\alpha_2}}\,,
\ee
\begin{equation}
\label{nr}
r_{\alpha \dot{\beta}} =dh_{\alpha\dot{\beta}} +\omega_\alpha{}^\gamma \wedge
h_{\gamma\dot{\beta}} +\bar{\omega}_{\dot{\beta}}{}^{\dot{\delta}}
\wedge h_{\alpha\dot{\delta}}\,.
\end{equation}
The zero-torsion condition $r_{\alpha\dot \beta}=0$
expresses the Lorentz connection $\omega $ and $\bar \omega $ via
derivatives of $h$. After that, the $\lambda -$independent part of the
curvature $2-$forms $R$ (\ref{nR}) and $\bar R $\,(\ref{nbR})
coincides with the
Riemann tensor.  Einstein equations imply that the Ricci tensor vanishes
up to a trace part proportional to the  cosmological constant.
This is equivalent to saying that only those
components of the tensors (\ref{nR}) and (\ref{nbR}) are allowed to be
non-vanishing which belong to the Weyl tensor.  As is well-known \cite{PR},
Weyl tensor is described by the fourth-rank mutually conjugated totally
symmetric multispinors $C_{\alpha _1\alpha _2\alpha _3\alpha _4}$ and $\bar
C_{\dot \alpha_1\dot \alpha_2\dot \alpha_3\dot \alpha_4}$.  Therefore, Einstein
equations with the cosmological term can be cast into the form
\begin{equation}
\label{e1} r_{\alpha \dot{\beta}} =0\,,
\end{equation}
\be
\label{e2}
R_{\alpha_1\alpha_2}=h^{\gamma_1 \dot{\delta}} \wedge h^{\gamma_2}{}_{\dot{\delta}}
C_{\alpha_1 \alpha_2 \gamma_1 \gamma_2}\,,
\ee
\be
\label{e3}
\bar{R}_{\dot{\beta}_1 \dot{\beta}_2}=h^{\eta\dot{\delta}_1}\wedge h_\eta{}^{\dot{\delta}_2}
\bar{C}_{\dot{\beta}_1 \dot{\beta}_2 \dot{\delta}_1 \dot{\delta}_2}\,.
\ee

It is convenient to think of the $0-$forms $C$ and $\bar C$ on the
right hand sides of (\ref{e2}) and (\ref{e3})
as of independent field variables which turn out to be equivalent to the Weyl
tensor due to the equations
 (\ref{e2}) and (\ref{e3})
themselves. {}From
 (\ref{e2}) and (\ref{e3})
it follows that the $0-$forms $C$ and $\bar C$ should obey certain differential
restrictions as a consequence of the Bianchi identities for the curvatures $R$
and $\bar R$. It is not difficult to make sure that these differential
restrictions can be equivalently rewritten in the form
\be
\label{b1}
D^L C_{\alpha_1 \alpha_2 \alpha_3 \alpha_4}=h^{\gamma\dot{\delta}}
C_{\alpha_1 \alpha_2 \alpha_3 \alpha_4 \gamma ,\dot{\delta}}\,,
\ee
\be
\label{b2}
D^L \bar{C}_{\dot{\beta}_1 \dot{\beta}_2 \dot{\beta}_3 \dot{\beta}_4}=h^{\gamma\dot{\delta}}
\bar{C}_{\gamma ,\,\dot{\beta}_1 \dot{\beta}_2 \dot{\beta}_3 \dot{\beta}_4
\dot{\delta}}\,,
 \ee
where $C_{\alpha_1\ldots\alpha_5,\dot\delta}$ and
$\bar C_{\gamma,\beta_1\ldots\beta_5}$
are new arbitrary multispinor field variables
which are totally symmetric in spinor indices of each type, while $D^L$ is
the Lorentz-covariant differential
\begin{equation}
D^L A_{\alpha\dot{\beta}}=dA_{\alpha\dot{\beta}}+\omega_\alpha{}^\gamma \wedge
A_{\gamma\dot{\beta}}+\bar{\omega}_{\dot{\beta}}{}^{\dot{\delta}} \wedge A_{\alpha\dot{\delta}}.
\end{equation}

Once again, Bianchi identities for the $l.h.s.'s$ of (\ref{b1}),
(\ref{b2}) impose certain
differential restrictions on
$C_{\alpha_1\ldots\alpha_5,\dot\delta}$
and
$\bar C_{\gamma,\beta_1\ldots\beta_5}$
which can be cast into the form analogous to (\ref{b1}), (\ref{b2})
 by virtue of
introducing new field variables
$C_{\alpha_1\ldots\alpha_6,\dot\delta_1\dot\delta_2}$
and
$\bar C_{\alpha_1\alpha_2,\dot\delta_1\ldots\dot\delta_6}$.
Continuation of this process leads to the following infinite chain of
differential relations:
\bee
\label{c2}
D^L C_{\alpha_1\ldots\alpha_{n+4},\,\dot{\beta}_1\ldots\dot{\beta}_n}=
h^{\gamma\dot{\delta}} C_{\alpha_1\ldots\alpha_{n+4}\gamma,\,\dot{\beta}_1\ldots\dot{\beta}_n\dot{\delta}}
- n(n+4)\lambda^2 h_{\{\alpha_1\dot{\beta}_1}
C_{\alpha_2\ldots\alpha_{n+4},\,\dot{\beta}_2\ldots\dot{\beta}_n\}_{\alpha,\dot{\beta}}}
+O(C^2)\,,
\eee
\bee
\label{bc2}
D^L\bar{C}_{\alpha_1\ldots\alpha_n,\,\dot{\beta}_1\ldots\dot{\beta}_{n+4}}=
h^{\gamma\dot{\delta}}\bar{C}_{\alpha_1\ldots\alpha_n\gamma,\,
\dot{\beta}_1\ldots\dot{\beta}_{n+4}
\dot{\delta}}
-n(n+4)\lambda^2 h_{\{\alpha_1\dot{\beta}_1}
\bar{C}_{\alpha_2\ldots\alpha_n,\,\dot{\beta}_2\ldots\dot{\beta}_{n+4}\}_{\alpha,\dot{\beta}}}
+O(C^2)\,.
\eee
All these relations contain no new dynamical information in addition to that
contained in the original Einstein equations in the form
(\ref{e1})-(\ref{e3}). Analogously
to the spin 0 case, (\ref{c2}) and (\ref{bc2}) merely express
highest $0-$forms
$C_{\alpha_1\ldots\alpha_{n+4},\dot\beta_1\ldots\dot\beta_n}$
and
$\bar C_{\alpha_1\ldots\alpha_{n},\dot\beta_1\ldots\dot\beta_{n+4}}$
via derivatives of the lowest $0-$forms
$\bar C_{\alpha_1\alpha_2\alpha_3\alpha_4}$
and
$\bar C_{\dot \beta_1\dot \beta_2\dot \beta_3\dot \beta_4}$
containing at the same time all consistency conditions for (\ref{e2}),
(\ref{e3})
and the equations (\ref{c2}), (\ref{bc2}) themselves. Thus, the system of
equations (\ref{e1})-(\ref{e3}), (\ref{c2}) and (\ref{bc2})
turns out to be dynamically
equivalent to the Einstein equations with the cosmological term.  It is this
form of the equations
which one arrives at in the analysis of the higher-spin
equations of the previous section in the spin$-2$ sector.
Let us note that although we know closed equations for higher-spins
(\ref{beq})-(\ref{eeq}),
for the case of pure gravity an
explicit form of all terms nonlinear in $C$ on the $r.h.s.'s$ of (\ref{c2})
and (\ref{bc2}) is not still known in all orders.
The form of $C^2-$ type terms was obtained in \cite{V6}.
The infinite set of the $0-$forms $C$ and $\bar C$
can be interpreted as a convenient basis in the
linear space of all on-mass-shell
nontrivial components of curvatures and their covariant derivatives.

The example of pure gravity can be generalized straightforwardly
to all higher spins as explained in \cite{V5}.
The general linearized equations read
\be
\label{s}
R_{\alpha_1\ldots \alpha_n \,,\gb_1 \ldots\gb_m}=
\delta (m)h^{\gamma_1 \dot{\delta}}
\wedge h^{\gamma_2}{}_{\dot{\delta}} C_{\alpha_1 \ldots \alpha_n \gamma_1
\gamma_2} +
\delta(n)h^{\eta\dot{\delta}_1}\wedge h_\eta{}^{\dot{\delta}_2}
\bar{C}_{\dot{\beta}_1 \ldots\dot{\beta}_m \dot{\delta}_1 \dot{\delta}_2}\,,
\ee
\bee
\label{ca}
D^L C_{\alpha_1\ldots\alpha_{n},\,\dot{\beta}_1\ldots\dot{\beta}_m}=
h^{\gamma\dot{\delta}}
C_{\alpha_1\ldots\alpha_{m}\gamma,\,\dot{\beta}_1\ldots\dot{\beta}_n\dot{\delta}}
{}&-& nm\lambda^2 h_{\{\alpha_1\dot{\beta}_1}
C_{\alpha_2\ldots\alpha_{m},\,
\dot{\beta}_2\ldots\dot{\beta}_n\}_{\alpha,\dot{\beta}}}
+O(C^2)\,.
\eee
Needless to say that it is this form
of the linearized equations which one derives from
the equation (\ref{dB}).

\section{Higher-Spin Equations of Motion in 2+1 Dimensions}

The situation for
$d=2+1$ is very much parallel to
that for the 3+1 dimensional case.
The full equations are again formulated in terms of the
generating functions $ W(z,y;\psi |x),$ $B(z,y;\psi |x)$ and $s(z,y;\psi |x)$
which depend on the space-time coordinates $x^\nu\,\, (\nu =0-2)$, auxiliary
commuting spinors $z_\ga $ and $y_\ga\,\, (\ga =1,2),$
and two Clifford elements $\psi^i
\,\,(i=1,2)$
\be
\{\psi^i ,\psi^j \} =2\delta^{ij}\,,
\ee
where $W=$ $dx^\nu W_\nu (z,y;\psi |x)$
and
  $ B=$ $B(z,y;\psi |x)$ are
space-time 1-form and  0-form,
respectively,
while  $ s=dz^\ga\,s_\ga (z,y;\psi |x)$ is a
space-time 0-form and $z-$space 1-form with the auxiliary anticommuting
differentials
\be
dz^\ga dz^\gb =-dz^\gb dz^\ga
\ee
which commute with
all other variables $z_\ga,\,\,y_\ga,\,\,\psi^i ,\,\,x^\nu$ and $ dx^\nu .\,$
Again, $s$
acts as a differential operator shifting
along $z\,-$directions and does not possess its
own degrees of freedom. It is expressed entirely (up to a pure gauge part) in
terms of the 0-form $B$ which serves as the generating function for
 matter
fields. The 1-form $W$ is the generating function for higher-spin
gauge fields which do not propagate in 2+1 dimensions but mediate
interactions of the matter fields.

We endow the space of functions $f(z,y)$ with the structure of
the star-product algebra by restricting the formula (\ref{star})
to the subspace of functions independent of the dotted spinors.

The equations which describe higher-spin interactions of massless
matter in 2+1 dimensions have the
form \cite{d3}
similar to that of the equations in 3+1 dimensions
\be
\label{d3}
dW=W*W\,,\qquad
dB=W*B-B*W \,,\qquad
ds=W*s-s*W\,,
\ee
\be
\label{ss3}
s*s=- dz_\alpha dz^\alpha (i+B)\,,
\ee
\be
\label{sb3}
s*B=-B*s \,.
\ee
These equations have ``unfolded form" and therefore
possess explicit general
coordinate invariance and gauge invariance under the
infinitesimal gauge transformations
\be
 \delta{W}=d\varepsilon -W*\varepsilon +\varepsilon *W\,,\quad
 \delta{B}=\varepsilon *B - B*\varepsilon\,,\quad
 \delta{s}=\varepsilon *s - s*\varepsilon \,,
\ee
where $\varepsilon  =\varepsilon  (z,y;\psi |x) $
is an arbitrary gauge parameter.

To make sure that the system (\ref{d3})-(\ref{sb3})
describes higher-spin interactions
of massless
matter fields in 2+1 dimensions, one has to analyze it in the linearized
approximation. This analysis is in many respects parallel to that carried out
 for the 3+1-dimensional case.

At the first stage, one fixes an appropriate vacuum solution. We assume
that $W$ and $s$ contain zero-order
nontrivial vacuum components $W_0 $ and $s_0 $
while $B$ starts from the first-order terms. Namely, we fix
\be
\label{v3}
B_0 =0\,, \quad
s_0 =dz^\ga z_\ga \,,\quad
W_0 =\frac{1}{4i} (\go _0{}^{\ga\gb} y_\ga y_\gb  +h_0{}^{\ga\gb}
y_\ga y_\gb \,\psi_1 )\,.
\ee

One observes that $s_0 $ acts as
$z-$ differential,
\be
[s_0, f]_* =-2i\partial f(z,y)\,,\qquad\partial =dz^\ga
\frac{\partial}{\partial z^\ga }
\ee
for every $f = f(z,y).\,$ As a result, the equations (\ref{v3}) solve the
equations (\ref{d3})-(\ref{ss3}) except for the
equations for $W$, which imposes additional
restrictions on $\go_0 $ and
$h_0 $
\bee
\label{nads3}
 d\omega_{0\,\alpha \beta}(x)&=&\omega_{0\,\alpha \gamma}(x)\wedge
\omega_{0\,\beta}{}^\gamma (x)+h_{0\,\alpha \delta}(x)\wedge
h_{0\,\beta}{}^{\delta}(x)\,,\\\nonumber
  dh_{0\,\alpha
\beta}(x)&=&\omega_{0\,\alpha \gamma}(x)\wedge
h_0{}^\gamma{}_{\beta}(x)+\omega_{0\,\beta \delta}(x)\wedge
h_{0\,\alpha}{}^{\delta}(x)\,.
\eee
According to the analysis of section 4
the fields $\go_{0\,\ga\gb}
$ and $h_{0\,\ga\gb} $ are identified with
 the background gravitational Lorentz connection and
dreibein, respectively.
It is worth mentioning
that it is the necessity to have a non-degenerate space-time
background metric that forces one to introduce
the non-vanishing background 1-form
$W_0$ since otherwise the equations (\ref{d3})-(\ref{sb3})
cannot be interpreted in terms of particles.

The explicit analysis then shows that the role of nontrivial dynamical
variables is played by the ``initial data"
$C(y;\psi |x)$=$B(0,y;\psi |x)$.
Expanding  $C$ as
\be
C(y;\psi |x)= C^{aux}(y;\psi_1 |x) +\psi_2 C^{mat}(y;\psi_1 |x)
\ee
one finds that $C^{mat}$ describes massless matter
fields while $C^{aux}$ describes some auxiliary fields which do not
carry dynamical degrees of freedom \cite{un}. The matter sector
contains two massless bosons described by the even functions
$C^{mat}(-y;\psi_1 |x)=$$C^{mat}(y;\psi_1 |x)$ and two
massless fermions
 described by the odd functions
$C^{mat}(-y;\psi_1 |x)=$$-C^{mat}(y;\psi_1 |x)$. The doubling is due to
 the dependence on the Clifford element $\psi_1$. Note that
in the bosonic sector there exists a reduction to a model
describing a single massless scalar. This reduction is not possible
however in presence of fermions.
Let us note that for
analogous reason we confined ourselves to the pure bosonic model
in the case of 3+1 theory. The full 3+1 dimensional theory which
involves fermions requires some additional variables analogous to
$\psi_i$ \cite{more} or matrix algebras analogous to those considered
in section 13.

The physical interpretation of the generating function
$C^{mat}(y;\psi_1 |x)$ is that the lowest modes of its
expansion in powers of $y$ are identified with the dynamical
fields, i.e.
$C^{mat}(0;\psi_1 |x)$ describes scalar while the linear part of
$C^{mat}(y;\psi_1 |x)$ describes spinor. The highest modes describe
all on--mass--shell nontrivial derivatives of the matter fields.
The important physical distinction of the 2+1 dimensional case from
the 3+1 dimensional one is that in the former case there is no room for the
Weyl tensors related to the gauge fields.
This is in accordance with the well
known fact that higher-spin gauge fields do not propagate in 2+1 dimensions.
In fact this
implies that the dynamics we analyze is
of the Chern-Simons type thus generalizing the pure higher-spin
dynamics of Blencowe to the case with non-trivial matter fields.

An interesting question which we cannot discuss in full
detail in this talk
is why the higher-spin equations have this particular form
in 3+1 and 2+1 dimensions. This mainly concerns the sector
of the constraints (\ref{sB}), (\ref{eeq}),
(\ref{ss3}) and (\ref{sb3}) since one can write a lot of other
 versions of invariant constraints. The point is that these constraints
are singled out by the requirement that the full theory must possess the
local Lorentz invariance in the physical sector of $z-$ independent
fields $C(y;\ldots )$. This property is not straightforward at all.
The reason is that the
vacuum solution like (\ref{v3}) is not invariant under the
Lorentz transformations rotating all spinors $z_\ga$ and $y_\ga$ because
higher-spin gauge transformations do not affect the differentials $dz_\ga$.
The requirement that Lorentz symmetry must act in the standard way
on the physical modes is necessary for
the proper relativistic field theory interpretation
of the model. This property can be shown to be
guaranteed by the constraints described
in this section and in section 9 but fail for different choices of
the constraints.

\section{D=2 Matter}

Let us now describe the new results on the
higher-spin interactions of d2 matter fields.

Originally it was observed in \cite{BB,WIN} that
one can construct Noether current interactions for a massless
scalar field in two dimensions in the form
\be
S=\half\int_{M^2} \Big( \partial_\nu \phi \partial^\nu \phi +
\sum_n gA^{\nu_1 \ldots \nu_{2n}}J_{\nu_1 \ldots \nu_{2n}} \Big )\,,
\ee
where $J_{\nu_1 \ldots \nu_{2n}}$ are some conserved currents
which have a form
\be
J_{\nu_1 \ldots \nu_{2n}}=\partial_{\nu_1}\ldots\partial_{\nu_n} (\phi)
\partial_{\nu_{n+1}}\ldots\partial_{\nu_{2n}} (\phi) +\ldots\,,
\ee
where dots denote some trace terms proportional to $\eta_{\nu_i \nu_j }$.
These currents generalize the usual stress tensor
\be
J_{\nu\mu}=T_{\nu\mu}=\partial_\nu \phi \partial_\mu \phi
-\half \Big (\partial_\rho \phi \Big )^2 \eta_{\nu\mu}
\ee
to higher spins. In the light-cone coordinates the currents
have only two on--mass--shell nontrivial components
\be
J_{\underbrace{+\ldots +}_{2n}}=\Big (\partial_+^n\phi \Big )^2\,,\qquad
J_{\underbrace{-\ldots -}_{2n}}=\Big (\partial_-^n\phi \Big )^2\,.
\ee
It was argued in \cite{BB} that these currents generate some
infinite-dimensional algebra later on called $W_{1+\infty}$.

Below we generalize
the models of
\cite{BB,WIN} by introducing gauge invariant interactions for
the higher-spin gauge fields so that no vanishing
current constraints on the matter fields are present in our model.
The model is
formulated in an explicitly
higher-spin gauge invariant and general coordinate invariant
fashion. A natural d2 background is AdS space-time.
The presented model is not conformal.
Since the full equations of motion have a form of some
zero-curvature equations and covariant constantness conditions without
any additional constraints, the model turns out to be
integrable. This unexpected property is specific for d=2
and allows us to formulate a simple
$BF-$ type action principle for the model.

Let us first reformulate free equations of motion for matter fields
in d2 AdS space
in the form of some covariant constantness conditions along the lines of
 the ``unfolded formulation'' described
in section 8.
Here we use light-cone coordinates
and consider the AdS background described by the
zweibein $h^\pm$ and Lorentz connection $\omega$ obeying the vacuum
equations
\be
\label{0c}
R^0 =d\omega +h^- \wedge h^+ =0\,,\qquad R^\pm =dh^\pm \pm 2\omega \wedge h^\pm
=0\,.
\ee

Consider the following system of equations
\be
\label{Cn}
D\phi_n =\alpha (n) h^- \phi_{n+2} +\beta (n) h^+ \phi_{n-2}\,,
\ee
where $D$ is the Lorentz covariant derivative,
\be
D\phi_n =d\phi_n +n \omega \phi_n\,.
\ee
This system is formally consistent (i.e. the Bianchi
identities are satisfied) provided that
the numerical parameters
$\alpha (n)$ and $\beta (n)$ obey the condition
\be
\label{ab}
\alpha (n)\beta (n+2) = \mu +\frac{1}{4}\, n(n+2)\,
\ee
and zero curvature conditions (\ref{0c})
are satisfied. Here $\mu$ is an arbitrary numerical
parameter. Note that the ambiguity in the coefficients $\alpha (n)$
and $\beta (n)$, which is not fixed by (\ref{ab}), is
irrelevant and reflects a
freedom in the rescaling  $\phi_n\rightarrow \gamma (n)\phi_n$.

To make sure that, e.g., the equations (\ref{Cn}) with even $n$
are equivalent
to the Klein-Gordon equation
let us introduce
the inverse zweibein $h^\nu_\pm$ and rewrite the system of equations
(\ref{Cn}) in the form
\be
\label{inv}
h^\nu_+ D_\nu \phi_n = \beta (n) \phi_{n-2}\,,\qquad
h^\nu_- D_\nu \phi_n = \alpha (n) \phi_{n+2}\,.
\ee
One observes that these equations
with $n=0$ express the fields $\phi_{\pm 2}$ in terms of
the first space-time derivatives of $\phi_{0}$.
Then the equations (\ref{inv}) with
$n=\pm 2 $
contain the Klein-Gordon equation and express the
fields $\phi_{\pm 4}$ via second space-time derivatives of
$\phi_{0}$. (Note that although the Klein-Gordon equation appears
twice, i.e. both in the first of the equations (\ref{inv}) with $n=2$ and
in the second one with $n=-2$, an appropriate combination of these equations
vanishes identically due to the Bianchi identities of the original equations
(\ref{Cn}) so that, effectively, the Klein-Gordon equation appears only
once.)
Finally, one finds that all higher-$n$ equations
in the system (\ref{inv}) either express the fields $\phi_m$ with $m\neq 0$
via higher derivatives of $\phi_0$ or encode all Bianchi identities for these
expressions imposing no additional dynamical conditions
on the field $\phi_0$. This analysis is parallel to that of
section 8.

As a result, the system
(\ref{Cn}) with even $n$ turns out to be dynamically equivalent to the
Klein-Gordon equation supplemented with some constraints which express all
highest $\phi_n$ via higher space-time derivatives of the dynamical field
$\phi_0$.  The situation with fermions ($n$ is odd) is analogous.
A physical meaning of the components $\phi_n$ is that
they describe all on-mass-shell nontrivial derivatives of the
dynamical boson and fermion fields,
generalizing the flat-space higher derivatives
$(\partial_+ )^n \phi$ and $(\partial_- )^n \phi$
to the AdS case.

By analogy with the analysis of the d=3+1 and d=2+1 cases one can
conjecture \cite{FL} that
the relevant higher-spin algebra is $shs(2)$, which gives rise to the set
of gauge fields
\be
\go_\nu =\sum_{n,m=0}^\infty \go_{\nu\,n,m} (\hy_+ )^n (\hy_- )^m
\ee
with the elementary oscillators obeying the relations
\be
[\hy_- ,\hy_+ ]=-2i\,.
\ee
This algebra contains the AdS  subalgebra spanned by the generators
\be
\label{gen} L^\pm = \frac{i}{4} (\hat{y}_\pm )^2\,,\qquad   L^0 = \frac{i}{4}
\{\hat{y}_+ ,\hat{y}_- \}
\ee
 obeying the $sp(2)$ commutation relations
\be
\label{com}
[L^0 , L^\pm
] = \pm 2 L^\pm\,, \qquad [L^- ,L^+ ]=L^0 \,.
\ee
Let us emphasize that since higher-spin gauge fields are not propagating
in 1+1 dimensions, at this stage a choice of the higher-spin algebra is
ambiguous enough. The only important property is that it should contain
$sp(2)$ as a subalgebra. A final choice can be done from the analysis
of interactions.

A less trivial problem is how to describe matter fields. From the linearized
analysis it follows that one has to introduce a one-parametric set $\phi_n$
with $-\infty < n< \infty$. Evidently, it does not work any longer
to take a function
$\Phi =\sum_{n,m=0}^\infty \Phi_{n,m}(\hy_+ )^n (\hy_- )^m$
as in $d=3,4$ since it
involves too many components. The idea to chose Fock
(i.e. metaplectic) representation
$|\Phi\rangle =\sum_{n=0}^\infty \Phi_{n}(a^\dagger )^n |0\rangle $
is not working either since it contains only a half of states.

The way out is tricky enough. One has to start with the tensor product of two
Fock spaces
\be
|\Phi (x)\rangle =
\sum_{n,m=0}^\infty \Phi_{n,m} (a^+ )^n (b^+ )^m |0\rangle
\ee
and then to gauge away all operators which contain $(a^+ b^+ )|\chi\rangle$
for any $\chi$. As a result one is left just with the appropriate set of
matter fields
\be
|\Phi (x)\rangle =\Big [
\sum_{n=1}^\infty \Big (\Phi_{n}^+ (x) (a^+ )^n +\Phi^-_n (x)(b^+ )^n
\Big ) +\Phi_0 (x) \Big ]|0\rangle\,.
\ee
In section 12 we show how this idea is realized for interacting
d2 matter fields.

\section{Higher-Spin-Matter Interactions in 1+1 Dimensions}

Analogously to the scheme developed for
d=3,4
the basic algebraic element is the associative algebra
$A$ of power series in the generating elements ${y}_\pm$ and
${z}_\pm$ endowed with the associative star-product
\begin{equation}
\label{weyl}
(f* g)(z,y)=(2\pi )^{-4} \int d^2 s d^2 t d^2 p d^2 q\, f(z+s,y+p) g(z-t,y+q)
exp[i(s_\alpha t^\alpha +p_\alpha q^\alpha )]\,
\end{equation}
such that
\be
\label{ha}
[{y}_- ,{y}_+ ]_* = -2i\,,\qquad [{z}_- ,{z}_+ ]_* = 2i\,,\qquad
[{y}_\alpha ,{z}_\beta ]_* = 0\,\qquad \alpha ,\,\beta =\pm\,.
\ee
It is also convenient to use the following equivalent set of variables
\be
\label{uv}
{u}_\alpha =\frac{1}{2}\,({z}_\alpha -{y}_\alpha )\,, \quad
{v}_\alpha =\frac{1}{2}\,({y}_\alpha +{z}_\alpha )\,,\quad
[{v}_\pm ,{u}_\mp ] =\mp i \,,\quad [{v}_\alpha ,{v}_\beta ] =
[{u}_\alpha ,{u}_\beta ]=0\,.
\ee

The important property of $A$ is that it contains an element $\Pi $
which is a projection operator, $\Pi^2 =\Pi$, and behaves
as a vacuum vector for the operators ${u}_\pm$ and ${v}_\pm$, i.e.
${v}_\pm * \Pi =0$ and $\Pi* {u}_\pm  =0$.  Its explicit
realization is
\be
\label{fock}
\Pi=\frac{1}{4}\, exp(i z_\alpha y^\alpha )\,.
\ee

To formulate the d2 higher-spin dynamics
it is useful to extend the algebra $A$ to ${\cal A}$
by virtue of the following
general procedure.
Given associative algebra $A$ and some projection operator $\Pi\in A$,
one defines the
algebra ${\cal A}$ such that its general element ${\bf a}\in {\cal A}$ is
equivalent to a set
of four elements of $A$,
${\bf a}$=$\{a,|a\rangle ,\langle a|,\langle a\rangle \}$,
obeying the properties
\be
\label{pr}
\{{\bf a}\in {\cal A}\vert\,\,\,
a,|a\rangle ,\langle a|,\langle a\rangle \in A\,;
\,\,|a\rangle \Pi =|a\rangle ,\,
\Pi\langle a| =\langle a| ,\,
\langle a \rangle \Pi = \Pi \langle a \rangle =\langle a \rangle\}\,.
\ee
The product law  $\circ$
in ${\cal A}$ is defined via the product law in $A$ as follows
\be
\label{fp}
{\bf a}\circ {\bf b} =\{ab + |a\rangle\langle b|\,, a |b\rangle
+|a\rangle \langle b \rangle \,, \langle a| b+\langle a \rangle\langle
b|\,,\langle a||b\rangle + \langle a \rangle\langle b \rangle \}\,.
\ee
This product law is associative. Note that the supertrace operation
$str_A$ in
$A$ induces the supertrace operation $str_{\cal A}$ in ${\cal A}$
\be
\label{str}
str_{\cal A} ({\bf a})=str_A (a+\langle a\rangle )\,,
\ee
where $str_A$ is the supertrace operation (\ref{nstr}).

In the context of the d2 dynamics this construction with the projection
operator (\ref{fock}) is used
to embed all matter and auxiliary fields into the adjoint
representation of ${\cal A}$.
Namely, to describe the higher-spin gauge interactions of d2 matter
fields we introduce the gauge one-form
${\bf W}(x|z_\alpha ,y_\alpha )$=
$dx^\nu{\bf W}_\nu (x|z_\alpha ,y_\alpha )$,
and the matter field zero-form ${\bf B}(x|z_\alpha ,y_\alpha )$,
both
in the adjoint representation of ${\cal A}$, i.e.
\be
{\bf W}
=\{W,\,|W\rangle ,\, \langle W| ,\,
 \langle W \rangle \},\qquad
{\bf B}
=\{B,\,|B\rangle ,\, \langle B| ,\,
 \langle B \rangle \}\,.
\ee

The full system of equations for interacting d2 matter fields has
a simple form of zero-curvature conditions:
\be
\label{1f}
{\bf R}\equiv d{\bf W} + {\bf W}\circ\wedge {\bf W} =0\,,\qquad
d{\bf B} + {\bf W}\circ
{\bf B}-{\bf B}\circ{\bf W} =0\,.
\ee
These equations can be derived from the B-F type action principle
\be
\label{action}
S= \int_{M_2} str_{\cal A}({\bf B}\,{\bf R})\,.
\ee

The model becomes dynamically non-trivial because
the 0-form
${\bf B}$ is supposed to have a nonvanishing vacuum value
\footnote{Note
that the physical vacuum values of the fields ${\bf B}$ and ${\bf W}$ have
nothing to do with the vacuum $\Pi $ of the algebra $A$ of auxiliary spinor
variables.}
\be
\label{vacb} {\bf B}_{vac} =
\{N,0,0,0\}\,,\qquad N={1\over 2i} z_- z_+ \,.
\ee
A physical
vacuum value of the gauge 1-form ${\bf W}$ is
\be
\label{vacw} {\bf W}_{vac} =
\{\omega^{gr},0,0,0\}\,,\qquad \omega^{gr} (x) = h^{+} (x) L^+ +h^{-} (x) L^- +
\omega (x) L^0\,,
\ee
where $L^\pm$ and $L^0$ are the $sl_2$ generators (\ref{gen}).

 One-forms $h^\pm (x)=dx^\nu h^\pm_\nu (x)$ and $\omega (x)=dx^\nu \omega_\nu
(x) $ describe inverse zweibein and Lorentz connection, respectively.  The
components of the gravitational field-strength two-form,

\be
R^{gr} = d\omega^{gr} +\omega^{gr}\wedge \omega^{gr}
= R^{+} L^+ +R^{-} L^- + R^0 L^0
\ee
identify, respectively, with the torsion tensor, $R^{+}$, $R^{-}$, and with the
Riemann tensor, $R^0$, shifted by a cosmological term $h^- \wedge h^+$.
The vacuum gravitational field is supposed to obey the
zero-curvature conditions (\ref{0c}) so that
the first of the equations (\ref{1f}) is satisfied. The second
one is also true because the vacuum value $N$ of ${\bf B}$ depends only
on  $z$ and therefore commutes with the background gravitational
field (\ref{vacw}) due to (\ref{ha}).

Higher-spin gauge fields correspond to higher-order terms of the
expansion of
${\bf W}(x|z_\alpha ,y_\alpha )$ in powers of the auxiliary spinor
variables.
The gauge connection ${\bf W}$ and the matter field ${\bf B}$ have
the standard transformation laws under the
higher-spin gauge transformations with the parameter
${\bf \xi} (x|z_\alpha ,y_\beta )$,
\be
\label{ng1}
\delta {\bf W} = d{\bf \xi} +{\bf W}\circ {\bf \xi}-{\bf \xi}\circ
{\bf W}\,,\qquad
\delta {\bf B} = {\bf B}\circ {\bf \xi}- {\bf \xi}\circ {\bf B}\,,
\ee
which leave invariant the equations (\ref{1f}) and the action (\ref{action}).
General coordinate invariance is explicit too.

A global symmetry subalgebra which acts
linearly on physical states is described by the parameters
 commuting with ${\bf B}_{vac}$,
\be
{\bf \xi}=(\xi_{vac}\,,0,0,\langle\xi_{vac}\rangle )
\ee
with an arbitrary Abelian parameter
$\langle\xi_{vac}\rangle$ and the parameter $\xi_{vac}$
of the form
\be
\label{vs}
\xi_{vac}=\sum_{n,m,k=0}^\infty \xi_{n,m,k} (x) N^k
(y_- )^n \,(y_+ )^m\,.
\ee
Since $N$ (\ref{vacb}) commutes with the oscillators $y_\pm$, the
generating elements of the $W_{1+\infty}$ algebra, one is left with the
non-negative part
of the loop extension  $\tilde{W}_{1+\infty}$  of $W_{1+\infty}$.

The topological form of the
action (\ref{action}) is analogous to the
topological form of the d2 gravitational action
discussed in \cite{gr} and to the higher-spin action proposed in \cite{FL}.
This analogy is not exact however because in the latter
models the zero-curvature equations are true
in absence of matter and do not
describe propagating degrees of freedom while the equations (\ref{1f}) are
shown below to describe interactions of propagating scalar and spinor fields,
which phenomenon turns out to be
possible because of using infinite multiplets of fields.

Another important point is that the non-vanishing vacuum value
of the zero-form ${\bf B}$ (\ref{vacb})
leads effectively to some
$W^2$ - type terms in the action that opens a way to a proper diagonalization of
the action at the linearized level. Practically, a problem of reducing the
quadratic part of the action (\ref{action}) to the standard form is
highly involved due to presence of infinitely many
auxiliary fields.

To analyze the equations (\ref{1f}) perturbatively one considers the fields of
the form ${\bf W}={\bf W}_{vac} + {\bf w}$ and ${\bf B}={\bf B}_{vac} + {\bf b}$
where ${\bf w}$ and ${\bf b}$ denote perturbations.  Propagating matter fields
belong to the mutually conjugated components $|b\rangle$ and $\langle b|$ of
${\bf b}$.  The linearized equations (\ref{1f}) in the sector
of the matter fields $|b\rangle$ read
\be
\label{le02}
d|b\rangle + w^{gr} |b\rangle =N |w \rangle \,.
\ee
This equation implies, first, that
$|w \rangle$ expresses via the matter fields $|b\rangle$ and, second, that it
imposes some differential equations on those components of the matter fields
which are not proportional to  $N$. Let us show that the latter
differential equations are just the equations for free matter
fields analyzed in section 11.

The linearized gauge transformation (\ref{ng1}) for the field
$|b\rangle$ takes the form
$\delta |b\rangle =  N |\xi \rangle +O({\bf b})$.
This implies that the field $|b\rangle$ contains some
Higgs part which can be gauged away and a
reminder which is to be shown to describe matter fields.
The standard Fock representation for
$|b\rangle$ is
$|b\rangle = b^l (u_+ ,u_-
) * \Pi$.
Since $N=\frac{1}{2i}z_+ z_-$=
$\frac{1}{2i}(u+v)_+ (u+v)_-$, the Higgs - type
component of the transformation law for $|b\rangle$ allows one to get rid of any
polynomial in $u_+ u_-$ in $b^l$. As a result one can chose a gauge
with respect to the transform (\ref{ng1}) with
\be \label{hg} b^l (u_+ ,u_- ) =
b^l_+ (u_+ ) +b^l_- (u_- ) +b^l_0\,,\qquad  b^l_+ (0 ) =b^l_- (0 )=0\,.
\ee
Fields of this form cannot be compensated by virtue of any transformation
(\ref{ng1}) and therefore can describe some dynamical degrees of freedom. By
expanding (\ref{hg}) in powers of $u_\pm$ one observes that,
in accordance with the idea sketched in the end of section 11,
the structure of
the gauge fixed mater field $b^l$ (\ref{hg}) is just of the form one expects for
d2 matter fields from (\ref{Cn}).

To work out explicit form of the field
equations one has to substitute (\ref{hg}) into (\ref{le02}), decompose the
left-hand-side of (\ref{le02}) into a part proportional to $N$, which is
compensated by an appropriate choice of $|\omega\rangle$, and a part depending
either only on $u_+$ or only on $u_-$ as in (\ref{hg}), which will impose some
equations on $b^l$.  Let us give the final result for the field equations and
the value of the field $|w\rangle$ =$ w^l (u_\pm ) *  \Pi$:
\begin{eqnarray}
\label{me}
D b^l &=&\frac{1}{4i} h^{+} \left( (u_+)^2 (b^l_+
+b^l_0 ) +i u_+ \dot{b}^l_- (0) -4\ddot{b}^l_- (u_- )+ \int_0^1
ds\,(3s+1)\ddot{b}^l_- (su_- ) \right)\nonumber\\ &+&\frac{1}{4i} h^{-} \left(
(u_-)^2 (b^l_- +b^l_0 ) -i u_- \dot{b}^l_+ (0) -4\ddot{b}^l_+ (u_+ )+ \int_0^1
ds\,(3s+1)\ddot{b}^l_+ (su_+ ) \right)\,,
\end{eqnarray}
\begin{eqnarray}
\label{wl} w^l (u_\pm )&=&-\omega b^l (u_\pm ) + \frac{i}{2}h^{+} \left(iu_+
\int^1_0 ds \dot{b}^l_- (su_- ) -\int_0^1 ds\,(2s+1)\ddot{b}^l_- (su_- )
\right)
\nonumber\\ &\phantom{=}&\phantom{\omega_0 b^l (\hat{u}_\pm )} +
\frac{i}{2}h^{-} \left(iu_- \int^1_0 ds \dot{b}^l_+ (su_+ ) +\int_0^1
ds\,(2s+1)\ddot{b}^l_+ (su_+ ) \right) \,,
\end{eqnarray}
where $\dot{f}(x)$= ${\partial\over \partial x} f(x)$ and $D$ is the Lorentz
covariant derivative.

One can check directly that the equation (\ref{me}) is formally
consistent thus corresponding to some particular
case of the equations (\ref{Cn}) with the coefficients of the form
(\ref{ab}).
By expanding the function $f^l$ into power series in either
$u_+$ or $u_-$
one finds that the coefficients indeed satisfy the condition (\ref{ab})
with $\mu =3/16$. This value is not occasional. It equals to the
value of the $sl_2$ Casimir operator for
the realization (\ref{vacw}).
There is a possibility to generalize
the proposed scheme to an arbitrary mass which we will discuss elsewhere
\cite{prep}.
Let us note that the parameter $\mu$
is measured here in units of the inverse radius
of the background AdS space-time and therefore tends to zero
in the flat limit.

Thus it is shown that the linearized equations for $|b\rangle$
describe properly linearized dynamics for d2 matter fields.
Analogously one can analyze the conjugate sector of $\langle b|$
to show that it describes conjugate matter fields.

An important property which
we do not prove explicitly here is that
all other components in ${\bf W}$ and ${\bf B}$ do not carry
their own degrees of freedom.
This can be shown for example with the aid of  the method
developed in \cite{un} where it was argued that any system of
covariant constantness equations for zero forms
cannot describe propagating modes if these zero forms
carry some finite-dimensional representations of the space-time
symmetry algebra which gives rise to the vacuum gravitational field.
Actually, in the model under consideration all components of the
zero form ${\bf B}$ contained in $B$ and $\langle B\rangle$
decompose into a sum of only
finite-dimensional representations of the AdS algebra under the adjoint action
of the generators (\ref{gen}).

Thus, the matter fields contained in $|b\rangle$ and the conjugated
fields $\langle b|$ are the only propagating degrees of freedom in the
system. All other fields are either auxiliary or
 mediate interactions of
the matter fields.
In particular this is the case for the
gravitational field which corresponds to the sector of the $W$ fields quadratic
in $y_\pm$ and for its higher-spin generalizations corresponding to higher powers
in $y_\pm$.
Due to the form of the product law (\ref{fp})
the matter fields contribute quadratically to the equations for the
gravitational field and its higher-spin analogs in agreement with what
one expects from the matter sources for the gravitational field.

The linearized analysis shows that the
Lorentz connection occurs only through the standard Lorentz
covariant derivative.
This is important and not completely trivial
property that can be shown to remain valid
in all orders in interactions \cite{prep}
and in fact fixes the form of the d2 higher-spin dynamics.

The remarkable property of the
proposed equations (\ref{1f}) is that having a form of some zero
curvature conditions they can be integrated explicitly at least
locally
\be
{\bf W}(x)= {\bf g}^{-1}(x)d{\bf g}(x)\,,\qquad
{\bf B}(x)= {\bf g}^{-1}(x){\bf B_0}{\bf g}(x)\,,
\ee
where ${\bf g}(x)$ is an arbitrary $x$-dependent invertible element
of ${\cal A}$ while ${\bf B_0}$ is an arbitrary $x$-independent element
of ${\cal A}$.
The novel feature compared to the theories in higher dimensions is
that there are no constraints on $B_0$.
Thus the presented non-linear model
turns out to be integrable due to
the specific form of the higher-spin interactions.

\section{Extended Higher-Spin Superalgebras}

An important property of the higher-spin equations in
all examples considered above is that
they remain consistent if all field variables $W$, $B$ and $s$
(the latter one in 3+1 and 2+1 dimensions) take values in
an arbitrary associative algebra $A$. The simplest possibility consists of
identifying $A$ with the matrix algebra, $Mat_N({\rm C})$, in which case
field variables possess additional matrix indices, $W \rightarrow  W_i{}^j$,
$B\rightarrow  B_i{}^j$ with $i,j = 1,...,N$.
This offers a way for constructing
higher-spin systems with nontrivial internal symmetries of Yang-Mills type.

One can address the question what are the most general truncated versions
of these extended higher-spin theories which still lead to consistent
higher-spin dynamics. A convenient way to classify consistent higher-spin
models is to analyze automorphisms of the higher-spin algebras which
leave invariant the dynamical equations.  In this way the problem was fully
analyzed for the 3+1 dimensional theory in \cite{KV} and can be analogously
analyzed for the cases of 2+1 and 1+1 models. To illustrate this issue let
us summarize here the basic results for the 3+1 dimensional model. Following
 \cite{KV} we consider only the vacuum higher-spin symmetries which
generalize the symmetry $hs(4)$ considered in section 5, i.e. those which
leave invariant the vacuum solutions of the full field equations and act
as true symmetries on the physical states.
We focus on the higher-spin algebras which lead to consistent
higher-spin systems with finite-dimensional internal symmetries.

It turns out that there exist three types of higher-spin algebras which reduce
to unitary, symplectic and orthogonal gauge algebras in the spin$-1$ Yang-Mills
sector. Unitary higher-spin algebras denoted $hu(n;m|4)$ can be realized as
$(n+m)\times (n+m)$ matrices with the elements depending on
operators $\hat y_\alpha $ and
$\hat{\bar{y}}_{\dot{\alpha}}$
(\ref{hei})
\begin{equation}
\label{P}
P_i{}^j(\hat{y},\hat{\bar{y}})={\kern 2em\raise 2.5ex\hbox{$n$}}{\kern -0.7em\raise -2.5ex\hbox{$m$}}\phantom{n}
\begin{tabular}{|l|l|}
\multicolumn{2}{l}{\kern 2.5em\hbox{$n$}\kern 6em\hbox{$m$}} \\
\hline
$P^E{}_{i^\prime}{}^{j^\prime}  (\hat{y},\,\hat{\bar{y}})\phantom{\biggl(}$  & $P^O{}_{i^\prime}{}^{j^{\prime\prime}}(\hat{y},\,\hat{\bar{y}})$ \\
\hline
$P^O{}_{i^{\prime\prime}}{}^{j^\prime}(\hat{y},\,\hat{\bar{y}})\phantom{\biggl(}$  & $P^E{}_{i^{\prime\prime}}{}^{j^{\prime\prime}}(\hat{y},\,\hat{\bar{y}})$ \\
\hline
\multicolumn{2}{c}{}
\end{tabular}
\end{equation}
 for the conditions that
\be
P^E{}_i{}^j (\hat{y},\hat{\bar{y}})=P^E{}_i{}^j(-\hat{y},-\hat{\bar{y}})\,,
\ee
\be
P^O{}_i{}^j (\hat{y},\hat{\bar{y}})=-P^O{}_i{}^j(-\hat{y},-\hat{\bar{y}})\,,
\ee
$i.e.$ the diagonal blocks $P^E$ are bosonic
(the power series coefficients carry even numbers of spinor
indices) while the off-diagonal blocks $P^O$ are fermionic
(the power series coefficients carry odd numbers
of spinor indices). In addition, it is assumed that
the elements of the matrices
$P$ obey the reality conditions
\begin{equation}
[P_i{}^j(\hat{y},\hat{\bar{y}})]^\dagger =-(i)^{\pi (P_i{}^j)}\,P_i{}^j(\hat{y},\hat{\bar{y}})
\end{equation}
 with
\begin{equation}
(\hat{y}_\alpha )^\dagger =\hat{\bar{y}}_{\dot{\alpha}},\qquad \pi (P^E
)=0,\qquad \pi (P^O )=1\,.
\end{equation}

The algebras $hu(n;m|4)$ contain higher-spin subalgebras of orthogonal and
symplectic types denoted $ho(n;m|4)$ and $husp(n;m|4)$,
 respectively, which also
give rise to consistent equations of motion for massless fields of all spins
via appropriate truncations of the higher-spin equations corresponding to
$hu(n;m|4)$ \cite{{V5},{KV}}. These subalgebras can be extracted from $hu(n;m|4)$ by
imposing the following conditions
\begin{equation}
P_k{}^l(\hat{y},\hat{\bar{y}}) =
-(i)^{\pi (P_k{}^l )}\,\eta^{lu} P_u{}^v(i\hat{y},i\hat{\bar{y}})\eta^{-1}{}_{vk}
\end{equation}
 with some nondegenerate bilinear form $\eta _{kl}$. If $\eta _{kl}$ is
symmetric, $\eta _{kl}=\eta _{lk}$, this leads to the orthogonal algebras
$ho(n;m|4)$. The skewsymmetric form, $\eta _{kl}=-\eta _{lk}$,
gives rise to the symplectic algebras $husp(n;m|4)$ ($n$ and $m$ should be even
for the latter case).

Spin$-1$ Yang-Mills subalgebras of the higher-spin algebras defined in this way
are spanned by the matrices independent of the operators $\hat y_\alpha $ and
$\hat{\bar y}_{\dot{\alpha}}$
As a result, the Yang-Mills subalgebras coincide with $u(n)\oplus u(m)$,
$o(n)\oplus o(m)$ and $usp(n)\oplus usp(m)$ for $hu(n;m|4)$, $ho(n;m|4)$ and
$husp(n;m|4)$, respectively. Thus, all types of compact
Lie algebras which belong to the
classical series $a_n$, $b_n$, $c_n$ and $d_n$
can be realized as spin-1 Yang-Mills
symmetries in appropriate higher-spin theories.

The multiplicities of massless spin$-s$ particles in higher-spin theories
based on extended superalgebras are \cite{KV}
\begin{equation}
\begin{tabular}{|l|c|c|c|} \hline \hfill spin                &
       $ \mbox{ odd }$            &   $\mbox{ even }  $                &
       $\mbox{ half-}$         \\ algebra\hfill                 &
&                                     &    $\mbox{- integer } $  \\ \hline

$hu(n;m|4)\phantom{\kern -1em\biggl(}$ &    $n^2 +m^2$ &    $n^2+m^2$ &    $\bar{n}\otimes m+n\otimes\bar{m}$ \\
\hline
$ho(n;m|4)\phantom{\kern -1em\biggl(}$ &   $\frac{1}{2}(n(n-1)+m(m-1))$ &    $\frac{1}{2}(n(n+1)+m(m+1))$ &   $n\otimes m$     \\
\hline
     $husp(n;m|4)\phantom{\kern -1em\biggl(}$ &   $\frac{1}{2}(n(n+1)+m(m+1))$ &    $\frac{1}{2}(n(n-1)+m(m-1))$ &   $n\otimes m$     \\
\hline
\end{tabular}
\end{equation}
Let us note that
the fields of all odd spins belong to the adjoint representations
of the corresponding Yang-Mills algebras while even spins always belong to
a reducible representation which contains a singlet component.
This is a highly
important property since such a singlet component corresponds to the spin$-2$
colorless field to be identified with graviton. In other words
the finite-dimensional algebra constituted by the elements
$(I\otimes$  $bilinears$ $ in$ $\hat y$ $, $ $\hat{\bar y})$
is a proper subalgebra of all higher-spin algebras.

Another important property is that the higher-spin superalgebras
are supersymmetric in the standard sense only if $n=m$. Indeed, one
observes that (averaged) numbers of bosons and fermions coincide only for this
case. Also, it can be easily verified that all higher-spin superalgebras with
$n=m$ contain the anti-de Sitter superalgebra $osp(1;4)$
 as a subalgebra.
An opposite case with $n=0$ or $m=0$ corresponds to
the purely bosonic higher-spin
theories. The simplest version of the higher-spin
action and equations of motion discussed in
sections 7 and 9 corresponds to the case of $hu(1;0|4)\sim hs(4).$

{}From the structure of the higher-spin
superalgebras it is clear
why consistent interactions for a spin $s\geq 2$
field in d=3+1 are only possible in presence of infinite sets of massless
fields of infinitely increasing spins.
The reason is that
any field of spin $s\geq 2$ corresponds to
generators of the algebra which are some
 $deg >2$ polynomials of $\hat{y}$ and $\hat{\bar{y}}$
so that their successive commutators lead to higher
and higher polynomials.
The same happens when one attempts to replace the unit matrix $I$ by some
non-Abelian matrix algebra for bilinear polynomials in $\hat y$ and
$\hat{\bar y}$,
$i.e$. to introduce spin$-2$ particles possessing a non-Abelian structure.

Let us note that the above construction can be generalized
to infinite-dimensional algebras $A$. In particular, one can consider the
higher-spin superalgebras $h\ldots(n;m|4)$ with $n \rightarrow  \infty $\,\,
or/and $\ m \rightarrow  \infty $,\, that will lead to theories with infinite
numbers of massless particles of every spin. Such theories can be of interest
in the context of spontaneous breakdown of higher-spin gauge symmetries
and a relationship with string theory.

\section{Concluding Remarks}

At present, the consistent dynamics
of higher-spin gauge theories is formulated in 3+1, 2+1 and 1+1
dimensions.
Higher-spin theories  generalize
quite naturally all conventional massless systems such as spin$-0$
Klein-Gordon field, spin$-1/2$ Weyl field, spin$-1$ Yang-Mills fields,
spin$-(3/2)2$ -(super) gravitational fields containing all of them as
subtheories.  The analysis of \cite{lop,mec} indicates that consistent
higher-spin interactions can be formulated in higher dimensions too.

In addition to the arguments in favor of the relationship
between higher-spin theories and string theory
mentioned in Introduction, there exists a curious
parallelism between the two types of theories
which consists of the observation that both
correspond to certain non-local objects. Actually, it is well-known that
nice properties of
strings originate from the fact that they
are linearly extended non-local objects.
Higher-spin gauge theories are much simpler
but still non-local in some sense. They correspond
to quantum-mechanically non-local point particles in the space of auxiliary
variables $\hat q=\hat y_1$ and $\hat p=\hat y_2$
which non-locality can be traced back to the non-locality of
the star-product (\ref{nweyl}).
The important question then is whether this quantum-mechanical non-locality
of the classical higher-spin theories is enough to improve quantum
behavior after they are quantized as field theories in d=3+1 space-time.

\bigskip
\noindent
{\bf Acknowledgments}
\bigskip

The research described in this report
was made possible
in part by Grant \# MQM000 from the International Science Foundation.
This work was supported in part by the Russian Basic Research
Foundation, grant
93-02-15541.

\end{document}